\documentclass[review]{elsarticle}

\usepackage{lineno,hyperref}
\modulolinenumbers[5]

\journal{Journal of \LaTeX\ Templates}

\usepackage{rotate,graphicx,url,color}
\usepackage{amssymb}	
\usepackage{amsthm}		
\usepackage{amsmath}	

\newcommand{\e}{\mathrm{I \! E}} 
\newcommand{\re}{\mathrm{I \! R}} 
\newcommand{\na}{\mathrm{I \! N}} 
\newcommand{\sinal}{\mathrm{sign}} 

\newcommand{\vx}{\textbf{x}}
\newcommand{\vz}{\textbf{z}}
\newcommand{\vtheta}{\pmb{\theta}}
\newcommand{\vlambda}{\pmb{\lambda}}

\begin{document}

\begin{frontmatter}

\title{Inference in a bimodal Birnbaum-Saunders model}

\author[mymainaddress]{Rodney V.\ Fonseca\corref{mycorrespondingauthor}}
\cortext[mycorrespondingauthor]{Corresponding author}
\ead{rvf1@de.ufpe.br}

\author[mymainaddress]{Francisco Cribari-Neto}

\address[mymainaddress]{Departamento de Estat\'{i}stica, Universidade Federal de Pernambuco, Cidade Universit\'{a}ria, Recife/PE, 50740-540, Brazil}

\begin{abstract}
We address the issue of performing inference on the parameters that index a bimodal extension of the Birnbaum-Saunders distribution ($\mathcal{BS}$). We show that maximum likelihood point estimation can be problematic since the standard nonlinear optimization algorithms may fail to converge. To deal with this problem, we penalize the log-likelihood function. The numerical evidence we present shows that maximum likelihood estimation based on such penalized function is made considerably more reliable. We also consider hypothesis testing inference based on the penalized log-likelihood function. In particular, we consider likelihood ratio, signed likelihood ratio, score and Wald tests. Bootstrap-based testing inference is also considered. We use a nonnested hypothesis test to distinguish between two bimodal $\mathcal{BS}$ laws. We derive analytical corrections to some tests. Monte Carlo simulation results and empirical applications are presented and discussed.
\end{abstract}


\end{frontmatter}

\linenumbers


\section{Introduction}\label{Se:Introducao}
The Birnbaum-Saunders distribution was proposed by \cite{birnbaum1969a} to model failure time due to fatigue under cyclic loading. In such a model, failure follows from the development and growth of a dominant crack. Based on that setup, the authors obtained the following distribution function:
\begin{equation}
\label{E:fda_bs}
F(x) = \Phi \left[ \frac{1}{\alpha}\left( \sqrt{\frac{x}{\beta}} - \sqrt{\frac{\beta}{x}} \right) \right], \quad x>0,
\end{equation}
where $\alpha>0$ and $\beta>0$ are shape and scale parameters, respectively, and $\Phi(\cdot)$ is the standard normal cumulative distribution function (CDF). We write $X\sim\mathcal{BS}(\alpha, \beta)$.

Maximum likelihood estimation of the parameters that index the $\mathcal{BS}$ distribution was first investigated by \cite{birnbaum1969b}. Bias-corrected estimators were obtained by \cite{lemonte2007improved} and \cite{lemonte2008bootstrap}. Improved maximum likelihood estimation of the $\mathcal{BS}$ parameters was developed by \cite{cysneiros2008birnbaum}. \cite{ng2003modified} compared the finite-sample performance of maximum likelihood estimators (MLEs) to that of estimators obtained using the modified method of moments. For details on the $\mathcal{BS}$ distribution, its main properties and applications, readers are referred to \cite{leiva2015birnbaum}.

Several extensions of the $\mathcal{BS}$ distribution have been proposed in the literature aiming at making the model more flexible. For instance, \cite{diaz2005new} and \cite{sanhueza2008generalized} used non-Gaussian kernels to extend the $\mathcal{BS}$ model. The $\mathcal{BS}$ distribution was also extended through the inclusion of additional parameters; see, e.g., \cite{diaz2006some}, \cite{owen2006new} and \cite{owen2015revisit}. More recently, extensions of the $\mathcal{BS}$ model were proposed by \cite{bourguignon2014new}, \cite{cordeiro2013extended}, \cite{cordeiro2014exponentiated}  and \cite{zhu2015birnbaum}. Alternative approaches are the use of scale-mixture of normals, as discussed by \cite{balakrishnan2009estimation} and \cite{patriota2012scale}, for example, and the use of mixtures of $\mathcal{BS}$ distributions, as in \cite{balakrishnan2011some}. Again, details can be found in \cite{leiva2015birnbaum}.

A bimodal $\mathcal{BS}$ distribution, which we denote by $\mathcal{BBS}$ distribution, was proposed by \cite{olmos2015}. The authors used the approach described in \cite{gomez2011bimodal} to obtain a variation of the $\mathcal{BS}$ model that can assume bimodal shapes. Another variant of the $\mathcal{BS}$ distribution that exhibits bimodality was discussed by \cite{diaz2006some} and \cite{owen2015revisit}, which the latter authors denoted by $\mathcal{GBS}_2$. In their model, bimodality takes place when two parameter values exceed certain thresholds. In what follows we shall work with the $\mathcal{BBS}$ model instead of the $\mathcal{GBS}_2$ distribution because in the former bimodality is controlled by a single parameter. Even though we shall focus on the $\mathcal{BBS}$ distribution, in some parts of the paper we shall consider the $\mathcal{GBS}_2$ law as an alternative model; see Section \ref{Sec:Teste_hip_nao_encaixados} for further details.

A problem with the $\mathcal{BBS}$ distribution we detected is that log-likelihood maximizations based on Newton or quasi-Newton methods oftentimes fail to converge. In this paper we analyze some possible solutions to such a problem, such as the use of resampling methods and the inclusion of a  penalization term in the log-likelihood function. 

\begin{figure}
\centering
\caption{Contour curves of the profile log-likelihood of $\alpha$ and $\gamma$ with $\beta=0.69$ (fixed) for the runoff amounts data. Panel (a) corresponds to no penalization and panel (b) follows from penalizing the log-likelihood function.}
\label{F:runoff_logliks}
\includegraphics[scale=.50]{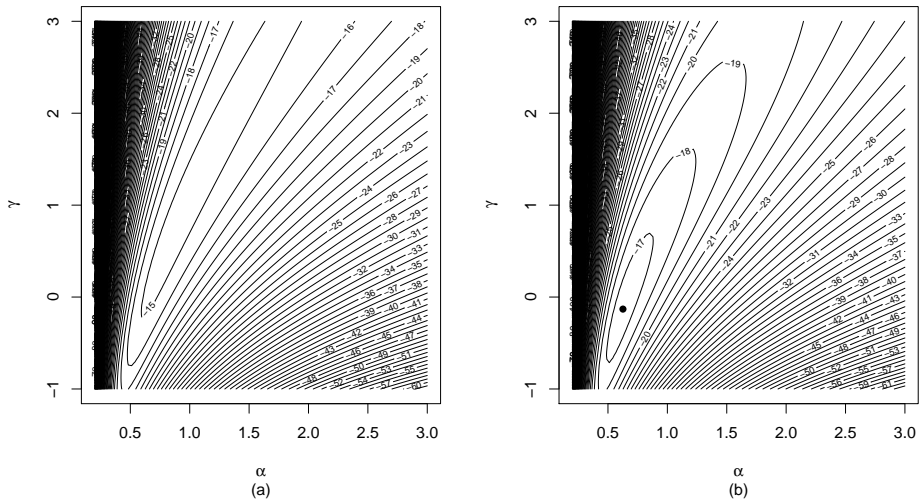}
\end{figure}

As a motivation, consider the data provided by \cite{folks1978inverse} that consist of 25 observations on runoff amounts at Jug Bridge, in Maryland. Figure~\ref{F:runoff_logliks}a shows log-likelihood contour curves obtained by varying the values of $\alpha$ and $\gamma$ while keeping the value of $\beta$ fixed. Notice that there is a region apparently flat of the profile log-likelihood function, which cause the optimization process to fail to converge. In Figure \ref{F:runoff_logliks}b we present similar contour curves for a penalized version of the log-likelihood function. It can be seen that plausible estimates are obtained. We shall return to this application in Section~\ref{Sec:aplicacao}.

The chief goal of our paper is to provide a solution to the convergence failure and implausible parameter estimates associated with log-likelihood maximization in the $\mathcal{BBS}$ model. We compare different estimation procedures and propose to include a penalization term in the log-likelihood function. In particular, regions of the parameter space where the likelihood is flat or nearly flat are heavily penalized. That approach considerably improves maximum likelihood parameter estimation. We also focus on hypothesis testing inference based on the penalized log-likelihood function. For instance, a one-sided hypothesis test is used to test whether the variate follows the $\mathcal{BBS}$ law with two modes. Analytical and bootstrap corrections are proposed to improve the finite sample performances of such test. Moreover, we present nonnested hypothesis tests that can be used to distinguish between two bimodal extensions of the $\mathcal{BS}$ distribution, the $\mathcal{BBS}$ and $\mathcal{GBS}_2$ models. The finite sample performances of all tests are numerically evaluated using Monte Carlo simulations.

The paper unfolds as follows. Section~\ref{Sec:dist_bbs} presents the $\mathcal{BBS}$ distribution and its main properties. Simulation results are presented in Section~\ref{Sec:func_veros}, where we outline some possible solutions to the numerical difficulties associated with $\mathcal{BBS}$ log-likelihood maximization. Two-sided hypothesis tests in the $\mathcal{BBS}$ model are discussed in Section~\ref{Sec:Teste_hip_bilaterais}. In Section~\ref{Sec:Teste_hip_unilaterais} we focus on one-sided tests where the main interest lies in detecting bimodality. Section~\ref{Sec:Teste_hip_nao_encaixados} describes nonnested hypothesis testing inference. Empirical applications are presented and discussed in Section~\ref{Sec:aplicacao}. Finally, some concluding remarks are offered in Section~\ref{Sec:Conclusao}.


\section{The bimodal Birnbaum-Saunders distribution}
\label{Sec:dist_bbs}

\begin{figure}
\centering
\caption{$\mathcal{BBS}(\alpha,\beta,\gamma)$ densities for some parameter values.}
\label{F:BBSpdf}
\includegraphics[width=\linewidth]{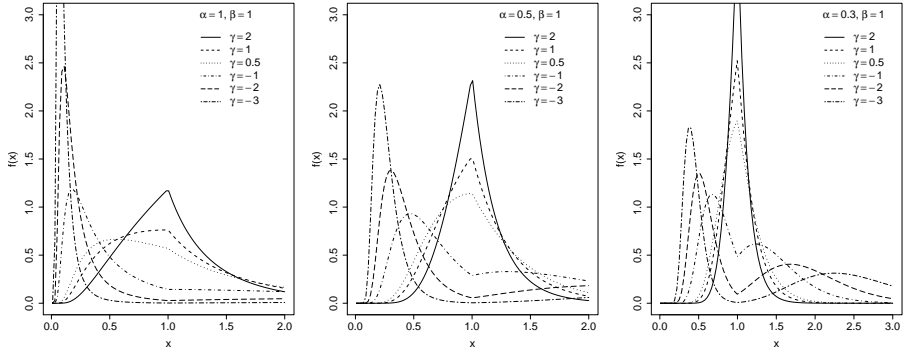}
\end{figure}

The Birnbaum-Saunders distribution proposed by \cite{olmos2015} can be used to model positive data and is more flexible than the original $\mathcal{BS}$ distribution since it can accommodate bimodality. A random variable $X$ is $\mathcal{BBS}$($\alpha$, $\beta$, $\gamma$) distributed if its probability density function (PDF) is given by 
\begin{equation}
f(x) = \frac{x^{-3/2}(x+\beta)}{4\alpha\beta^{1/2}\Phi(-\gamma)}\phi(|t|+\gamma), \quad x>0,
\label{E:fdp_bbs}
\end{equation}
where $\alpha,\beta>0$, $\gamma \in \re$, $t = \alpha^{-1}(\sqrt{x/\beta}-\sqrt{\beta/x})$ and $\phi(\cdot)$ is the standard normal PDF. Figure~\ref{F:BBSpdf} shows plots of the density in (\ref{E:fdp_bbs}) for some parameter values. We note that when $\gamma < 0$ the density is bimodal.

The CDF of $X$ is 
\begin{equation}
F(x) =  \left[ \frac{\Phi(t-\gamma)}{2\Phi(-\gamma)} \right]^{I(x,\beta)}
\left[\frac{1}{2}+\frac{\Phi(t-\gamma)-\Phi(\gamma)}{2\Phi(-\gamma)} \right]^{1-I(x,\beta)}, \quad x>0, 
\label{E:fda_bbs}
\end{equation}
where 
\begin{equation} 
I(x,\beta)=\left\{\begin{array}{cc}
1 & \text{if } x<\beta, \\
0 & \text{if } x\geq \beta.
\end{array} \right.
\end{equation} 
Some key properties of the $\mathcal{BS}$ distribution also hold for the $\mathcal{BBS}$ model, such as proportionality and reciprocity closure, i.e., $aX \sim \mathcal{BBS}(\alpha,a\beta,\gamma)$ and $X^{-1}\sim\mathcal{BBS}\left(\alpha,\beta^{-1},\gamma\right)$, respectively, where $a$ is a positive scalar.

An expression for the $r$th ordinary moment of $X$ is
\begin{equation}
\label{E:momento_r}
\e(X^r) = \frac{\beta^r}{\Phi(-\gamma)}\sum_{k=0}^{r}\sum_{j=0}^{k}\sum_{s=0}^{m}\binom{2r}{2k}\binom{k}{j}
\binom{m}{s}\left(\frac{\alpha}{2}\right)^m(-\gamma)^{m-s}d_s(\gamma),
\end{equation}
where $r\in\na$ and $d_a(r)$ is the $r$th standard normal incomplete moment: 
$$
d_r(a) = \int_{a}^{\infty} t^r\phi(t)dt.
$$

A useful stochastic representation is $Y = |T| + \gamma$. Here, $Y$ follows the truncated standard normal distribution with support in $(\gamma,\infty)$, $T = (\sqrt{X/\beta} - \sqrt{\beta/X})/\alpha$ and $X \sim \mathcal{BBS}(\alpha,\beta,\gamma)$. This relationship can be used to compute moments of the $\mathcal{BBS}$ distribution.


\section{Log-likelihood functions}
\label{Sec:func_veros}

Consider a row vector $\vx = (x_1, \ldots, x_n)$ of independent and identically distributed (IID) observations from the $\mathcal{BBS}(\alpha, \beta, \gamma)$ distribution. Let $\vtheta = (\alpha,\beta,\gamma)$ be the vector of unknown parameters to be estimated. The log-likelihood function is
\begin{small}
\begin{align}
\ell(\vtheta)  &= -n\log\left\{ 4\alpha\beta^{1/2}\Phi(-\gamma)(2\pi)^{1/2} \right\} -\frac{3}{2}\sum_{i=1}^{n}\log(x_i) + \sum_{i=1}^{n}\log(x_i + \beta) \nonumber \\ 
&- \frac{1}{2}\sum_{i=1}^{n}(|t_i| +\gamma)^2.
\label{E:loglik}
\end{align}
\end{small}
Differentiating the log-likelihood function with respect to each parameter we obtain the score function $U_{\vtheta} = (U_{\alpha},U_{\beta},U_{\gamma})$, where
\begin{small}
\begin{align}
U_{\alpha} = \frac{\partial \ell(\vtheta)}{\partial\alpha} &= -\frac{n}{\alpha} + \frac{1}{2}\sum_{i=1}^{n}t_i^2 + \frac{\gamma}{\alpha}\sum_{i=1}^{n}|t_i|, \label{E:loglik_der_a}\\
U_{\beta} = \frac{\partial \ell(\vtheta)}{\partial \beta} &= -\frac{n}{2\beta} + \sum_{i=1}^{n}\frac{1}{x_i+\beta} + \sum_{i=1}^{n}\frac{\sinal(t_i)(|t_i|+\gamma)}{2\alpha\beta^{3/2}}\left( x_i^{1/2} + \frac{\beta}{x_i^{1/2}} \right), \label{E:loglik_der_b} \\
U_{\gamma} = \frac{\partial \ell(\vtheta)}{\partial \gamma} &= n\frac{\phi(\gamma)}{\Phi(-\gamma)} - n\gamma - \sum_{i=1}^{n}|t_i|  \label{E:loglik_der_g},
\end{align}
\end{small}
and $\sinal(\cdot)$ represents the sign function.

The parameters MLEs, namely $\hat{\vtheta} = (\hat{\alpha}, \hat{\beta}, \hat{\gamma})$, can be obtained by solving $U_{\vtheta}=0$. They cannot be expressed in closed-form and parameter estimates are obtained by numerically maximizing the log-likelihood function using a Newton or quasi-Newton algorithm. To that end, one must specify an initial point for the iterative scheme. We propose using as starting values for $\alpha$ and $\beta$ their modified method of moments estimates \citep{ng2003modified}, and also using $\gamma=0$ as a starting value; the latter means that the algorithm starts at the $\mathcal{BS}$ law. We used such starting values in the numerical evaluations, and they proved to work well.

Based on several numerical experiments we noted a serious shortcoming: iterative numerical maximization of the $\mathcal{BBS}$ log-likelihood function may fail to converge and may yield implausible parameter estimates. Indeed, that is very likely to happen, especially when $\gamma > 0$. It is not uncommon for one to obtain very large (thus implausible) $\mathcal{BBS}$ parameter estimates, which is indicative that the likelihood function may be monotone; see \cite{pianto2011dealing}. We shall address this problem in the subsections that follow. 

\subsection{Log-likelihood function penalized by the Jeffreys prior} 

An interesting estimation procedure was proposed by \cite{firth1993bias}, where the score function is modified in order to reduce the bias of the MLE. An advantage of this method is that maximum likelihood estimates need not be finite since the correction is applied in a preventive fashion. For models in the canonical exponential family, the correction can be applied directly to likelihood function:
$$
L^*(\vtheta|\vx) = L(\vtheta|\vx)|K|^{1/2},
$$
where $|K|$ is the determinant of the expected information matrix. Thus, penalization of the likelihood function entails multiplying the likelihood function by the Jeffreys invariant prior.

Even though the $\mathcal{BBS}$ distribution is not a member of the canonical exponential family, we shall consider the above penalization scheme. In doing so, we follow \cite{pianto2011dealing} who used the same approach in speckled imagery analysis. We seek to prevent cases of monotone likelihood function that might lead to frequent optimization nonconvergences and implausible estimates. The $\mathcal{BBS}(\alpha, \beta, \gamma)$ expected information matrix was obtained by \cite{olmos2015}. Its determinant is
$$
|K| = \left[ L_{\beta\beta} + \frac{1}{\alpha^2\beta^2} + \frac{\gamma(\gamma - \omega)}{4\beta^2} \right]\left[ \frac{(\gamma - \omega)\omega(3-\gamma \omega - \gamma^2) + 2}{\alpha^2} \right],
$$
where $\omega = \phi(\gamma)/\Phi(-\gamma)$ and $L_{\beta\beta} = \e\left[(X+\beta)^{-2}\right]$. Thus, the log-likelihood function penalized by the Jeffreys prior can be written as
\begin{align}
\ell^*(\vtheta)  &= -n\log\left\{ 4\alpha\beta^{1/2}\Phi(-\gamma)(2\pi)^{1/2} \right\} -\frac{3}{2}\sum_{i=1}^{n}\log(x_i) + \sum_{i=1}^{n}\log(x_i + \beta) \nonumber \\ 
&- \frac{1}{2}\sum_{i=1}^{n}(t_i^2 + 2|t_i|\gamma +\gamma^2) + \frac{1}{2}\log\left[ L_{\beta\beta} + \frac{1}{\alpha^2\beta^2} + \frac{\gamma(\gamma - \omega)}{4\beta^2} \right]  \nonumber \\
&+ \frac{1}{2}\log\left[ \frac{(\gamma - \omega)\omega(3+\gamma(\gamma - \omega)) + 2}{\alpha^2} \right].
\label{E:loglik_Jeff_pen}
\end{align}  

If the likelihood function is monotone, the function becomes very flat for large parameter values and the Jeffreys penalization described above essentially eliminates such parameter range from the estimation. The likelihood of nonconvergences taking place and implausible estimates being obtained should be greatly reduced.  

\subsection{Log-likelihood function modified by the better bootstrap}

An alternative approach uses the method proposed by \cite{cribari2002improved}, where bootstrap samples are used to improve maximum likelihood estimation similarly to the approach introduced by \cite{efron1990more} and known as `the better bootstrap'. The former, however, does not require the estimators to have closed-form expressions. Based on the sample $\vx = (x_1, \ldots, x_n)$ of $n$ observations, we obtain pseudo-samples $\vx^*$ of the same size by sampling from $\vx$ with replacement. Let $P_i^*$ denote the proportion of times that observation $x_i$ is selected, $i=1,\ldots,n$. We obtain the row vector $\mathbf{P}^{*b} = (P_1^{*b}, \ldots, P_n^{*b})$ for the $b$th pseudo-sample, $b=1,\ldots,B$. Now compute a row vector $\mathbf{P}^*(\cdot)$ as
\begin{equation*}
\mathbf{P}^*(\cdot) = \frac{1}{B}\sum_{b=1}^{B}P^{*b},
\end{equation*}
i.e., compute the vector of mean selection frequencies using the $B$ bootstrap samples. The vector $\mathbf{P}^*(\cdot)$ is then used to modify the log-likelihood function in the following manner:
\begin{align}
\ell(\vtheta)  &= -n\log\left\{ 4\alpha\beta^{1/2}\Phi(-\gamma)(2\pi)^{1/2} \right\} -\frac{3n}{2}\mathbf{P}^*(\cdot)\log(\vx)^{\top} + n\mathbf{P}^*(\cdot)\log(\vx + \beta)^{\top} \nonumber \\ 
&- \frac{n}{2}\mathbf{P}^*(\cdot)\mathbf{t}_{\gamma}^{\top},
\label{E:loglik_Bboot}
\end{align}
where $\log(\vx) = (\log(x_1), \ldots, \log(x_ n))$, $\log(\vx + \beta)=(\log(x_1$ $+ \beta)$,$\ldots$,$\log(x_n + \beta))$ and $\mathbf{t}_{\gamma} = ((|t_1|+\gamma)^2, \ldots, (|t_n|+\gamma)^2)$ are row vectors. Hence, $\mathbf{P}^*(\cdot)$ is used to obtain a weighted average of the log-likelihood function terms that involve the data. The motivation behind the method is to approximate the ideal bootstrap estimates (which corresponds to $B=\infty$) faster than with the usual nonparametric bootstrap approach. In this paper we shall investigate whether this method is able to attenuate the numerical difficulties associated with $\mathcal{BBS}$ log-likelihood function maximization.  

\subsection{Log-likelihood function with a modified Jeffreys prior penalization}

Monotone likelihood cases can arise with considerable frequency in models based on the asymmetric normal distribution, with some samples leading to situations where maximum likelihood estimates of the asymmetry parameter may not be finite, as noted by \cite{liseo1990classe}. A solution to such a problem was proposed by \cite{sartori2006bias}, who used the score function transformation proposed by \cite{firth1993bias} in the asymmetric normal and Student-$t$ models. A more general solution was proposed by \cite{azzalini2013maximum}, who penalized the log-likelihood function as follows:  
$$
\ell^{*}(\vtheta) = \ell(\vtheta) - Q,
$$
where $\ell(\vtheta)$ and $\ell^{*}(\vtheta)$ denote the log-likelihood function and its modified version, respectively. The authors imposed some restrictions on $Q$, namely: (i) $Q \geq 0$; (ii) $Q=0$ when the asymmetry parameter equals zero (values close to zero can lead to monotone likelihood cases in the asymmetric normal model); (iii) $Q\rightarrow \infty$ when the asymmetry parameter in absolute value tends to infinity. Additionally, $Q$ should not depend on the data or, at least, be $O_p(1)$. According to \cite{azzalini2013maximum}, when these conditions are satisfied, the estimators obtained using $\ell^{*}(\vtheta)$ are finite and have the same asymptotic properties as standard MLEs, such as consistency and asymptotic normality.

We shall now use a similar approach for the $\mathcal{BBS}$ model. In particular, we propose modifying the Jeffreys penalization term so that the new penalization satisfies the conditions listed by \cite{azzalini2013maximum}. Since the numerical problems are mainly associated with $\alpha$ and $\gamma$, only terms involving these parameters were used. We then arrive at the following penalization term: 
\begin{equation}
\label{E:penalizacao_Q}
Q = Q_{\gamma} + Q_{\alpha} 
= -\frac{1}{2}\log\left\{ \frac{(\gamma-\omega)\omega[3+\gamma(\gamma-\omega)]}{2} + 1 \right\} + \frac{1}{2}\log(1+\alpha^2),
\end{equation}
where, as before, $\omega = \phi(\gamma)/\Phi(-\gamma)$. 

\begin{figure}
\centering
\caption{$Q_{\alpha}$ and $Q_{\gamma}$, modified Jeffreys penalization.}
\label{F:EMV_penal_QaQg}
\includegraphics[width=\linewidth]{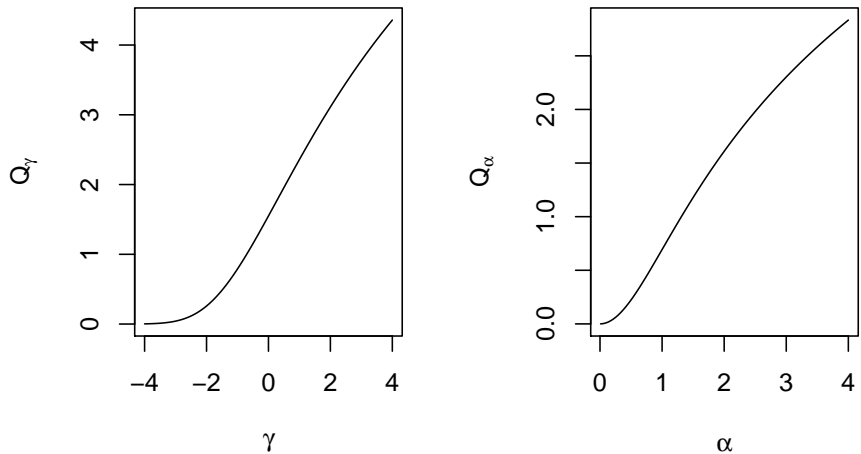}
\end{figure}

We note that $Q_{\alpha} \geq 0$. Additionally, $Q_{\alpha} \rightarrow 0$ when $\alpha \rightarrow 0$, and $Q_{\alpha} \rightarrow \infty$ when $\alpha \rightarrow \infty$. It can be shown that $Q_{\gamma} \rightarrow \infty$ when $\gamma \rightarrow \infty$ and that $Q_{\gamma} \rightarrow 0$ when $\gamma \rightarrow -\infty$, such that $Q_{\gamma} \geq 0$. Figure~\ref{F:EMV_penal_QaQg} shows the penalization terms as a function of the corresponding parameters. The quantities $Q_{\gamma}$ and $Q_{\alpha}$ penalize large positive values of $\gamma$ and $\alpha$, helping avoid estimates that are unexpectedly large. Therefore, the proposed penalization satisfies the conditions indicated by \cite{azzalini2013maximum}. An advantage of the penalization scheme we propose is that, unlike the Jeffreys penalization, it does not require the computation of $L_{\beta\beta}$. In what follows we shall numerically evaluate the effectiveness of the proposed correction when performing point estimation and we shall also consider the issue of carrying out testing inference on the parameters that index the $\mathcal{BBS}$ model.

\subsection{Numerical evaluation}

A numerical evaluation of the methods described in this section was performed. We considered different $\mathcal{BBS}$ estimation strategies. In what follows we shall focus on the estimation of the bimodality parameter $\gamma$.

\begin{figure}
\centering
\caption{Nonconvergence proportions (pnf) for different estimation methods using different values of $\gamma$.}
\label{F:EMV_nfs}
\includegraphics[scale=.50]{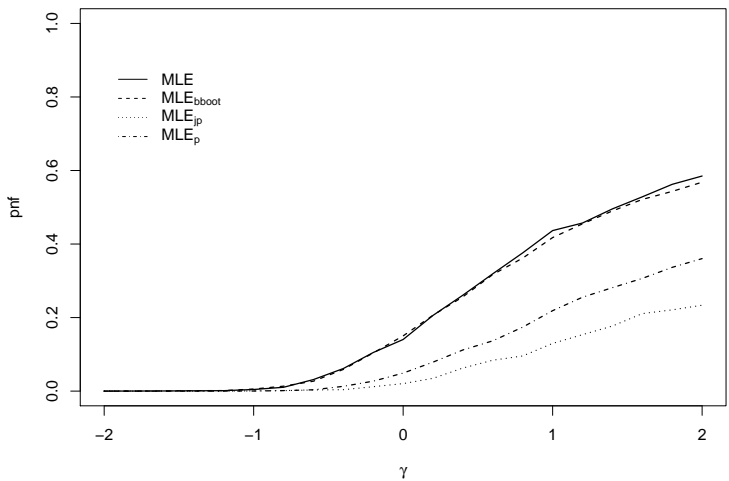}
\end{figure}

The Monte Carlo simulations were carried out using the \verb|Ox| matrix programming language \citep{doornik2009object}. Numerical maximizations were performed using the BFGS quasi-Newton method. We considered alternative nonlinear optimization algorithms such as Newton-Raphson and Fisher's scoring, but they did not outperform the BFGS algorithm. We then decided to employ the BFGS method, which is typically regarded as the best performing method \citep[Section 8.13]{mittelhammer2000econometric}. The results are based on 5,000 Monte Carlo replications for values of $\gamma$ ranging from $-2$ to $2$ and samples of size $n=50$. In each replication, maximum likelihood estimates were computed and it was verified whether the nonlinear optimization algorithm converged. At the end of the experiment, the frequency of nonconvergences (proportion of samples for which there was no convergence) was computed for each method (denoted by pnf). Figure~\ref{F:EMV_nfs} shows the proportion of nonconvergences corresponding to the standard MLEs, the MLEs obtained using the better bootstrap ($\text{MLE}_{\text{bboot}}$) and the MLEs obtained from the log-likelihood function penalized using the Jeffreys prior ($\text{MLE}_{\text{jp}}$) and its modified version ($\text{MLE}_{\text{p}}$) as a function of $\gamma$. Notice that MLE and $\text{MLE}_{\text{bboot}}$ are the worst performers when $\gamma > 0$; they display the largest rates of nonconvergence. The methods based on penalized log-likelihood function display the smallest values of pnf, with slight advantage for $\text{MLE}_{\text{jp}}$. 

\begin{figure}
\centering
\caption{Nonconvergence proportions (pnf) for different estimation methods using different sample sizes.}
\label{F:EMV_nfs_variando_n}
\includegraphics[scale=.50]{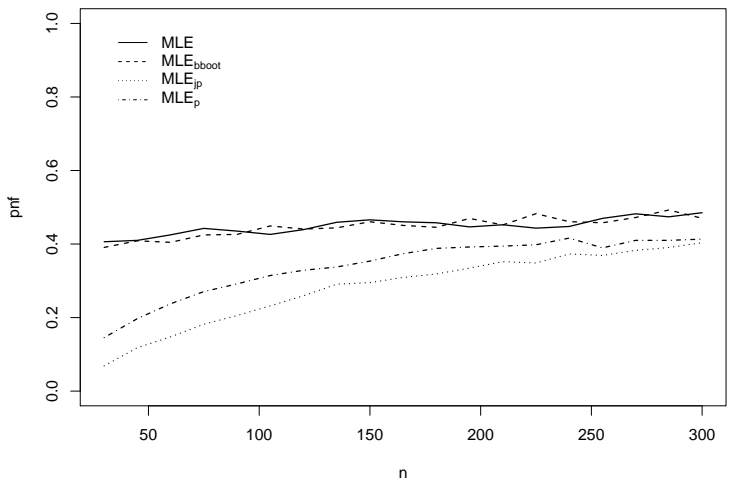}
\end{figure}

In order to evaluate the impact of the sample size on nonconvergence rates, a numerical study similar to the previous one was performed, but now with the value of the bimodality parameter fixed at $\gamma=1$. The samples sizes are $n\in \{30,45,60,75,\ldots,300\}$. The number of Monte Carlo replications was 5,000 for each value of $n$. The results are displayed in Figure~\ref{F:EMV_nfs_variando_n}. We note that the sample size does not seem to influence the MLE and $\text{MLE}_{\text{bboot}}$ nonconvergence rates. The corresponding optimizations failed in approximately 40\% of the samples regardless of the sample size. In contrast, the $\text{MLE}_{\text{jp}}$ and $\text{MLE}_{\text{p}}$ failure rates display a slight increase and then stabilize as $n$ increases. Recall that one of the conditions imposed by \cite{azzalini2013maximum} on the penalization term is that it should remain $O_p(1)$ as $n\rightarrow \infty$, i.e., the penalization influence seems to decrease as larger sample sizes are used, which leads to slightly larger nonconvergence frequencies in larger samples.

A second set of Monte Carlo simulations was carried out, this time only considering the estimator that uses the better bootstrap resampling scheme and also estimators based on the two penalized likelihood functions, i.e., we now only consider  $\text{MLE}_{\text{bboot}}$, $\text{MLE}_{\text{jp}}$ and $\text{MLE}_{\text{p}}$. Again, 5,000 Monte Carlo replications were performed. We estimated the bias (denoted by B) and mean squared errors (denoted by MSE) of the three estimators. The number of nonconvergences is denoted by nf. Tables~\ref{T:emv_pj}, \ref{T:emv_p} and \ref{T:emv_bboot} contain the results for $\text{MLE}_{\text{jp}}$, $\text{MLE}_{\text{p}}$ and $\text{MLE}_{\text{bboot}}$, respectively. Overall, $\text{MLE}_{\text{p}}$ outperforms $\text{MLE}_{\text{jp}}$. For instance, when $n=30$ in the last combination of parameter values, the MSEs of $\hat{\alpha}_{\text{jp}}$, $\hat{\beta}_{\text{jp}}$ and $\hat{\gamma}_{\text{jp}}$ are, respectively, 0.0211, 0.0022 and 2.6671, whereas the corresponding values for $\hat{\alpha}_{\text{p}}$, $\hat{\beta}_{\text{p}}$ and $\hat{\gamma}_{\text{p}}$ are 0.0167, 0.0020 and 2.0471. $\text{MLE}_{\text{bboot}}$ is typically less biased when it comes to the estimation of $\alpha$ and $\gamma$, but there are more convergence failures when computing better bootstrap estimates. Overall, the estimator based on the log-likelihood function that uses the penalization term we proposed typically yields more accurate estimates than $\text{MLE}_{\text{jp}}$ and outperforms $\text{MLE}_{\text{bboot}}$ in terms of convergence rates.

\begin{table}[htp]
\centering
\caption{Bias and mean squared error of $\text{MLE}_{\text{jp}}$ for some combinations of parameter values.}
\label{T:emv_pj}
\begin{small}
\begin{tabular}{cccccccc}
\hline
$n$ & $\widehat{\mathrm{B}}(\hat{\alpha}_{\text{jp}})$ & $\widehat{\text{B}}(\hat{\beta}_{\text{jp}})$ & $\widehat{\text{B}}(\hat{\gamma}_{\text{jp}})$ & $\widehat{\text{MSE}}(\hat{\alpha}_{\text{jp}})$  & $\widehat{\text{MSE}}(\hat{\beta}_{\text{jp}})$ & $\widehat{\text{MSE}}(\hat{\gamma}_{\text{jp}})$ & nf \\
 \hline
\multicolumn{8}{c}{$\alpha=0.5$, $\beta=1$ and $\gamma=-1$} \\
 \hline
 30 & $-0.0851$ & $-0.0048$ & $-0.4369$ & 0.0134 & 0.0118 & 0.3770 &   1 \\
 50 & $-0.0567$ & $-0.0032$ & $-0.2844$ & 0.0079 & 0.0068 & 0.2022 &   7 \\
100 & $-0.0303$ & $-0.0025$ & $-0.1498$ & 0.0038 & 0.0032 & 0.0876 &   2 \\
150 & $-0.0208$ & $-0.0016$ & $-0.1018$ & 0.0025 & 0.0021 & 0.0574 &   0 \\
 \hline
\multicolumn{8}{c}{$\alpha=0.5$, $\beta=1$ and $\gamma=0$} \\
 \hline
 30 & $-0.1532$ & $-0.0037$ & $-0.8487$ & 0.0303 & 0.0103 & 0.9668 &  67 \\
 50 & $-0.1103$ & $-0.0017$ & $-0.5798$ & 0.0184 & 0.0057 & 0.5201 & 124 \\
100 & $-0.0692$ & $-0.0002$ & $-0.3523$ & 0.0095 & 0.0026 & 0.2476 & 185 \\
150 & $-0.0484$ & $-0.0006$ & $-0.2412$ & 0.0061 & 0.0017 & 0.1510 & 181 \\
 \hline
\multicolumn{8}{c}{$\alpha=0.5$, $\beta=1$ and $\gamma=1$} \\
 \hline  
 30 & $-0.2295$ & $0.0001$ & $-1.5358$ & 0.0585 & 0.0064 & 2.6802 & 373 \\
 50 & $-0.1865$ & $-0.0018$ & $-1.1795$ & 0.0403 & 0.0031 & 1.6360 & 729 \\
100 & $-0.1332$ & $-0.0009$ & $-0.8061$ & 0.0229 & 0.0012 & 0.8366 & 1520 \\
150 & $-0.1070$ & $-0.0005$ & $-0.6337$ & 0.0164 & 0.0008 & 0.5793 & 2115 \\
 \hline
\multicolumn{8}{c}{$\alpha=0.3$, $\beta=1$ and $\gamma=-1$} \\
 \hline
 30 & $-0.0520$ & $0.0004$ & $-0.4429$ & 0.0048 & 0.0050 & 0.3812 &   5 \\
 50 & $-0.0353$ & $-0.0005$ & $-0.2874$ & 0.0029 & 0.0028 & 0.2068 &   2 \\
100 & $-0.0174$ & $-0.0013$ & $-0.1400$ & 0.0013 & 0.0013 & 0.0875 &   2 \\
150 & $-0.0119$ & $-0.0003$ & $-0.0970$ & 0.0009 & 0.0009 & 0.0578 &   1 \\
 \hline
\multicolumn{8}{c}{$\alpha=0.3$, $\beta=1$ and $\gamma=0$} \\
 \hline
 30 & $-0.0914$ & $-0.0023$ & $-0.8477$ & 0.0109 & 0.0039 & 0.9736 &  86 \\
 50 & $-0.0663$ & $-0.0018$ & $-0.5873$ & 0.0067 & 0.0022 & 0.5309 & 127 \\
100 & $-0.0401$ & $-0.0003$ & $-0.3406$ & 0.0033 & 0.0010 & 0.2366 & 205 \\
150 & $-0.0298$ & $0.0000$ & $-0.2472$ & 0.0022 & 0.0006 & 0.1512 & 208 \\
 \hline
\multicolumn{8}{c}{$\alpha=0.3$, $\beta=1$ and $\gamma=1$} \\
 \hline
 30 & $-0.1380$ & $0.0002$ & $-1.5331$ & 0.0211 & 0.0022 & 2.6671 & 419 \\
 50 & $-0.1114$ & $0.0001$ & $-1.1799$ & 0.0144 & 0.0011 & 1.6382 & 745 \\
100 & $-0.0808$ & $-0.0009$ & $-0.8149$ & 0.0084 & 0.0004 & 0.8538 & 1652 \\
150 & $-0.0649$ & $0.0000$ & $-0.6450$ & 0.0059 & 0.0003 & 0.5863 & 2211 \\
\hline
\end{tabular}
\end{small}
\end{table}

\begin{table}[htp]
\centering
\caption{Bias and mean squared error of $\text{MLE}_{\text{p}}$ for some combinations of parameter values.}
\label{T:emv_p}
\begin{small}
\begin{tabular}{cccccccc}
\hline
$n$ & $\widehat{\mathrm{B}}(\hat{\alpha}_{\text{p}})$ & $\widehat{\text{B}}(\hat{\beta}_{\text{p}})$ & $\widehat{\text{B}}(\hat{\gamma}_{\text{p}})$ & $\widehat{\text{MSE}}(\hat{\alpha}_{\text{p}})$  & $\widehat{\text{MSE}}(\hat{\beta}_{\text{p}})$ & $\widehat{\text{MSE}}(\hat{\gamma}_{\text{p}})$ & nf \\
 \hline
\multicolumn{8}{c}{$\alpha=0.5$, $\beta=1$ and $\gamma=-1$} \\
 \hline
 30 & $-0.0575$ & $0.0078$ & $-0.3129$ & 0.0114 & 0.0121 & 0.3151 &  12 \\
 50 & $-0.0364$ & $0.0006$ & $-0.1918$ & 0.0072 & 0.0068 & 0.1730 &   4 \\
100 & $-0.0190$ & $0.0018$ & $-0.0999$ & 0.0035 & 0.0032 & 0.0795 &   0 \\
150 & $-0.0128$ & $0.0004$ & $-0.0670$ & 0.0025 & 0.0021 & 0.0546 &   0 \\
 \hline
\multicolumn{8}{c}{$\alpha=0.5$, $\beta=1$ and $\gamma=0$} \\
 \hline
 30 & $-0.1163$ & $0.0035$ & $-0.6441$ & 0.0236 & 0.0100 & 0.7143 & 214 \\
 50 & $-0.0804$ & $0.0042$ & $-0.4358$ & 0.0151 & 0.0056 & 0.4199 & 240 \\
100 & $-0.0477$ & $0.0015$ & $-0.2505$ & 0.0082 & 0.0026 & 0.2069 & 287 \\
150 & $-0.0323$ & $0.0009$ & $-0.1697$ & 0.0057 & 0.0017 & 0.1369 & 318 \\
 \hline
\multicolumn{8}{c}{$\alpha=0.5$, $\beta=1$ and $\gamma=1$} \\
 \hline  
 30 & $-0.1965$ & $0.0035$ & $-1.3012$ & 0.0474 & 0.0057 & 2.0986 & 857 \\
 50 & $-0.1535$ & $0.0002$ & $-0.9703$ & 0.0321 & 0.0029 & 1.2726 & 1398 \\
100 & $-0.1058$ & $0.0003$ & $-0.6446$ & 0.0185 & 0.0011 & 0.6702 & 2200 \\
150 & $-0.0815$ & $0.0011$ & $-0.4877$ & 0.0136 & 0.0007 & 0.4739 & 2688 \\
 \hline
\multicolumn{8}{c}{$\alpha=0.3$, $\beta=1$ and $\gamma=-1$} \\
 \hline
 30 & $-0.0326$ & $0.0012$ & $-0.2940$ & 0.0042 & 0.0049 & 0.3119 &  14 \\
 50 & $-0.0197$ & $0.0005$ & $-0.1725$ & 0.0025 & 0.0029 & 0.1672 &   7 \\
100 & $-0.0102$ & $0.0005$ & $-0.0909$ & 0.0013 & 0.0013 & 0.0846 &   1 \\
150 & $-0.0065$ & $0.0002$ & $-0.0590$ & 0.0009 & 0.0009 & 0.0556 &   0 \\
 \hline
\multicolumn{8}{c}{$\alpha=0.3$, $\beta=1$ and $\gamma=0$} \\
 \hline
 30 & $-0.0682$ & $0.0012$ & $-0.6436$ & 0.0085 & 0.0036 & 0.7375 & 252 \\
 50 & $-0.0474$ & $0.0020$ & $-0.4288$ & 0.0053 & 0.0021 & 0.4085 & 298 \\
100 & $-0.0277$ & $0.0012$ & $-0.2424$ & 0.0029 & 0.0009 & 0.2034 & 349 \\
150 & $-0.0190$ & $0.0005$ & $-0.1686$ & 0.0020 & 0.0006 & 0.1340 & 349 \\
 \hline
\multicolumn{8}{c}{$\alpha=0.3$, $\beta=1$ and $\gamma=1$} \\
 \hline
 30 & $-0.1161$ & $0.0005$ & $-1.2810$ & 0.0167 & 0.0020 & 2.0471 & 979 \\
 50 & $-0.0920$ & $-0.0005$ & $-0.9769$ & 0.0116 & 0.0010 & 1.2924 & 1475 \\
100 & $-0.0624$ & $0.0000$ & $-0.6381$ & 0.0067 & 0.0004 & 0.6761 & 2491 \\
150 & $-0.0478$ & $0.0005$ & $-0.4844$ & 0.0049 & 0.0003 & 0.4685 & 3066 \\
\hline
\end{tabular}
\end{small}
\end{table}

\begin{table}[htp]
\centering
\caption{Bias and mean squared error of $\text{MLE}_{\text{bboot}}$ for some combinations of parameter values.}
\label{T:emv_bboot}
\begin{small}
\begin{tabular}{cccccccc}
\hline
$n$ & $\widehat{\mathrm{B}}(\hat{\alpha}_{\text{bboot}})$ & $\widehat{\text{B}}(\hat{\beta}_{\text{bboot}})$ & $\widehat{\text{B}}(\hat{\gamma}_{\text{bboot}})$ & $\widehat{\text{MSE}}(\hat{\alpha}_{\text{bboot}})$  & $\widehat{\text{MSE}}(\hat{\beta}_{\text{bboot}})$ & $\widehat{\text{MSE}}(\hat{\gamma}_{\text{bboot}})$ & nf \\
 \hline
\multicolumn{8}{c}{$\alpha=0.5$, $\beta=1$ and $\gamma=-1$} \\
 \hline
 30 & $-0.0283$ & $0.0091$ & $-0.1575$ & 0.0141 & 0.0121 & 0.3221 & 126 \\
 50 & $-0.0130$ & $0.0035$ & $-0.0761$ & 0.0086 & 0.0070 & 0.1898 &  38 \\
100 & $-0.0075$ & $0.0017$ & $-0.0430$ & 0.0040 & 0.0033 & 0.0861 &   2 \\
150 & $-0.0041$ & $0.0018$ & $-0.0230$ & 0.0027 & 0.0021 & 0.0568 &   0 \\
 \hline
\multicolumn{8}{c}{$\alpha=0.5$, $\beta=1$ and $\gamma=0$} \\
 \hline
 30 & $-0.0811$ & $0.0056$ & $-0.4596$ & 0.0281 & 0.0093 & 0.7186 & 1494 \\
 50 & $-0.0476$ & $0.0038$ & $-0.2658$ & 0.0177 & 0.0054 & 0.4071 & 1224 \\
100 & $-0.0215$ & $0.0001$ & $-0.1192$ & 0.0093 & 0.0025 & 0.2052 & 860 \\
150 & $-0.0138$ & $0.0001$ & $-0.0780$ & 0.0061 & 0.0016 & 0.1365 & 657 \\
 \hline
\multicolumn{8}{c}{$\alpha=0.5$, $\beta=1$ and $\gamma=1$} \\
 \hline  
 30 & $-0.1609$ & $0.0030$ & $-1.0607$ & 0.0498 & 0.0053 & 1.9644 & 4640 \\
 50 & $-0.1203$ & $0.0026$ & $-0.7657$ & 0.0380 & 0.0028 & 1.2973 & 5096 \\
100 & $-0.0735$ & $0.0003$ & $-0.4606$ & 0.0228 & 0.0012 & 0.7213 & 5877 \\
150 & $-0.0562$ & $0.0001$ & $-0.3442$ & 0.0154 & 0.0007 & 0.4782 & 5926 \\
 \hline
\multicolumn{8}{c}{$\alpha=0.3$, $\beta=1$ and $\gamma=-1$} \\
 \hline
 30 & $-0.0159$ & $0.0016$ & $-0.1554$ & 0.0050 & 0.0049 & 0.3204 & 150 \\
 50 & $-0.0090$ & $0.0028$ & $-0.0820$ & 0.0031 & 0.0029 & 0.1927 &  43 \\
100 & $-0.0040$ & $0.0008$ & $-0.0384$ & 0.0015 & 0.0013 & 0.0883 &   2 \\
150 & $-0.0032$ & $0.0008$ & $-0.0292$ & 0.0009 & 0.0008 & 0.0551 &   0 \\
 \hline
\multicolumn{8}{c}{$\alpha=0.3$, $\beta=1$ and $\gamma=0$} \\
 \hline
 30 & $-0.0472$ & $0.0024$ & $-0.4434$ & 0.0101 & 0.0034 & 0.7141 & 1543 \\
 50 & $-0.0280$ & $-0.0000$ & $-0.2633$ & 0.0070 & 0.0020 & 0.4315 & 1252 \\
100 & $-0.0127$ & $0.0003$ & $-0.1150$ & 0.0034 & 0.0009 & 0.2104 & 813 \\
150 & $-0.0090$ & $0.0007$ & $-0.0823$ & 0.0023 & 0.0006 & 0.1416 & 698 \\
 \hline
\multicolumn{8}{c}{$\alpha=0.3$, $\beta=1$ and $\gamma=1$} \\
 \hline
 30 & $-0.0983$ & $0.0007$ & $-1.0801$ & 0.0179 & 0.0019 & 1.9779 & 4528 \\
 50 & $-0.0716$ & $0.0002$ & $-0.7586$ & 0.0131 & 0.0010 & 1.2703 & 5128 \\
100 & $-0.0428$ & $0.0004$ & $-0.4492$ & 0.0094 & 0.0004 & 0.7788 & 5844 \\
150 & $-0.0328$ & $-0.0000$ & $-0.3410$ & 0.0059 & 0.0003 & 0.5082 & 6142 \\
\hline
\end{tabular}
\end{small}
\end{table}

\begin{figure}
\centering
\caption{Mean squared errors of the estimators of $\alpha$ (a), $\beta$ (b) and $\gamma$ (c), and the number of nonconvergences nf (d), for different values of $\gamma$ and $\phi$.}
\label{F:EQMs_nf_phi}
\includegraphics[scale=.50]{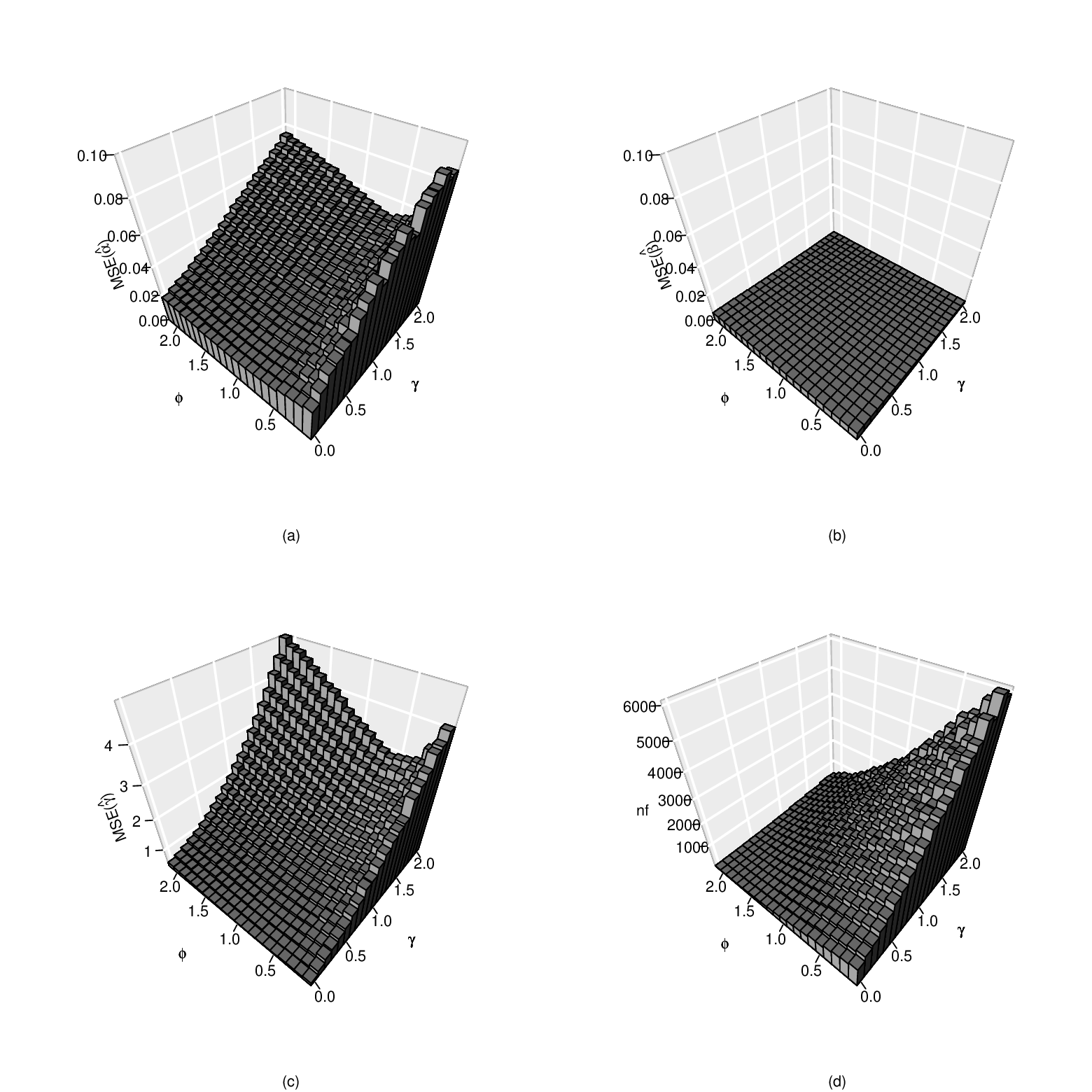}
\end{figure}

Next, we shall evaluate how changes in the penalization term impact the frequency of nonconvergences when computing $\text{MLE}_{\text{p}}$. In particular, we consider the following penalized log-likelihood function: 
$$
\ell^{*}_{\phi}(\vtheta) = \ell(\vtheta) - Q^{\phi},
$$
with $\phi>0$ fixed. This additional quantity controls for the penalization strength, with $\phi=1$ resulting in $\text{MLE}_{\text{p}}$ and different values of $\phi$ leading to stronger or weaker penalizations. A Monte Carlo study was performed to evaluate the accuracy of the parameter estimates for $\gamma \in \{0,0.1,\ldots,2.0 \}$ and $\phi \in \{ 0.1, 0.2, \ldots, 2.1 \}$. The parameter values are $\alpha = 0.5$ and $\beta=1$, and the sample size is $n=50$. Again, 5,000 replications were performed for each combination of values of $\phi$ and $\gamma$. Samples for which there was  convergence failure were discarded. Figure~\ref{F:EQMs_nf_phi} shows the estimated MSEs and the number of nonconvergences for each combination of $\gamma$ and $\phi$. Figures~\ref{F:EQMs_nf_phi}a and \ref{F:EQMs_nf_phi}b show that estimates of $\alpha$ are less accurate than those of $\beta$,  both being considerably more accurate than the estimates of $\gamma$ (Figure~\ref{F:EQMs_nf_phi}c). The MSE of the estimator of $\gamma$ tends to be smaller when the value of $\phi$ is between $0.4$ and 1, especially for larger values of $\gamma$. Visual inspection of Figure~\ref{F:EQMs_nf_phi}d shows that larger values of $\gamma$ lead to more nonconvergences, which was expected in light of our previous results. Furthermore, nf tends to decrease when larger values of $\phi$ are used. 

\begin{figure}
\centering
\caption{Number of nonconvergences (solid line) and MSE of $\hat{\gamma}$ (dashed line) for different values of $\gamma$ with $\phi\in\{0.5,1.0,1.5\}$. The nf values are shown in the left vertical axis and the values of $\widehat{\text{MSE}}(\hat{\gamma})$ are shown in the right vertical axis.}
\label{F:EQMg_nf_phi}
\includegraphics[scale=.40]{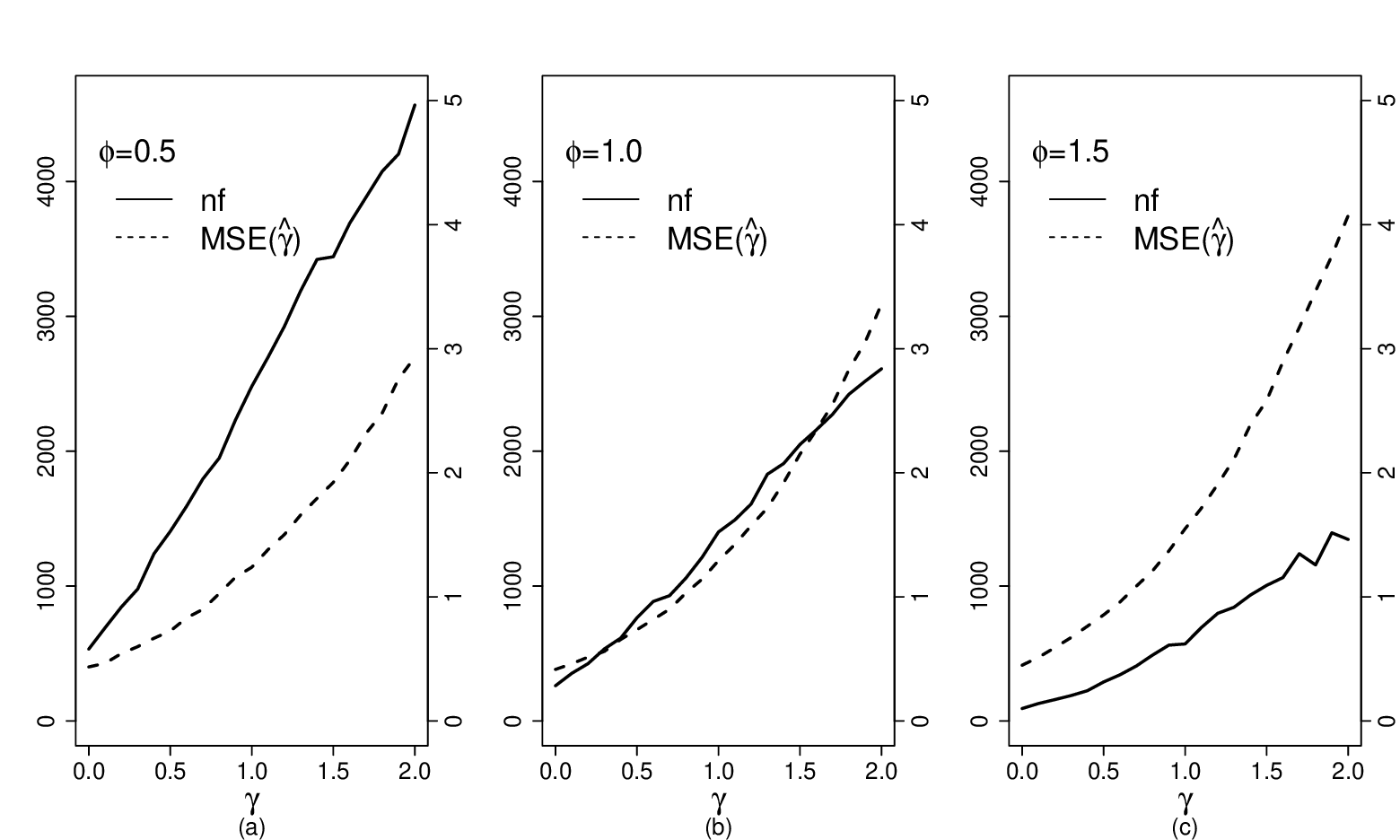}
\end{figure}

We note from Figure~\ref{F:EQMs_nf_phi} that the estimates of $\gamma$ are the ones most sensitive to changes in the values of $\gamma$ and $\phi$. Figure~\ref{F:EQMg_nf_phi} presents the number of nonconvergences (right vertical axis) and the MSE of $\hat{\gamma}$ (left vertical axis) as a function of $\gamma$ for three different values of $\phi$. Figure~\ref{F:EQMg_nf_phi}a shows that although $\phi=0.5$ yields more accurate estimates it also leads to more nonconvergences. For $\phi=1.5$, the number of nonconvergences did not exceed 1400, but $\widehat{\text{MSE}}(\hat{\gamma})$ was larger relative to other values of $\phi$. Overall, $\phi=1$ seems to balance well accuracy and the likelihood of convergence. In what follows we shall use $\phi=1$. 

 
\section{Two-sided hypothesis tests}
\label{Sec:Teste_hip_bilaterais}

In this section we consider two-sided hypothesis tests in the $\mathcal{BBS}$ model. Our interest lies in investigating the finite-sample performances of tests based on $\text{MLE}_{\text{p}}$. The first test we consider is the penalized likelihood ratio test, denoted by LR. Consider a model parametrized by $\vtheta=(\psi, \vlambda)$, where $\psi$ is the parameter of interest and $\vlambda$ is a nuisance parameter vector. Our interest lies in testing $H_0:\psi=\psi_0$ against a two-sided alternative hypothesis. The LR test statistic is
$$
W = 2\{\ell^{*}(\hat{\vtheta}) - \ell^{*}(\tilde{\vtheta}) \},
$$
where $\hat{\vtheta}$ is the unrestricted $\text{MLE}_{\text{p}}$ of $\vtheta$, i.e., $\hat{\vtheta}$ is obtained by maximizing $\ell^{*}(\vtheta)$ without imposing restrictions on the parameters, and $\tilde{\vtheta}$ is the restricted $\text{MLE}_{\text{p}}$, which follows from the maximization of $\ell^{*}(\vtheta)$ subject to the restrictions in $H_0$. Critical values at the $\epsilon\times100\%$ significance level are obtained from the null distribution of $W$ which, based on the results in \cite{azzalini2013maximum}, can be approximated by $\chi^2_1$ when $\psi$ is scalar. When $\psi$ is a vector of dimension $q$ ($\leq 3$), the test is performed in similar fashion with the single difference that the critical value is obtained from $\chi^2_q$.

It is also possible to test $H_0:\psi=\psi_0$ against $H_1:\psi\neq\psi_0$ using the score and Wald tests. To that end, we use the score function and the expected information matrix obtained using the penalized log-likelihood function. The score and Wald test statistics are given, respectively, by
\begin{align*}
W_{S} &= U^{*}(\tilde{\vtheta})^{\top}K^{*}(\tilde{\vtheta})^{-1}U^{*}(\tilde{\vtheta}),\\
W_{W} &= (\hat{\psi}-\psi_0)^2/K^{*}(\hat{\vtheta})^{\psi\psi},
\end{align*}
where $U^{*}(\vtheta)$ and $K^{*}(\vtheta)$ denote the score function and the expected information, respectively, obtained using the penalized log-likelihood function and $K^{*}(\vtheta)^{\psi\psi}$ is the diagonal element of the inverse of $K^{*}(\vtheta)$ corresponding to $\psi$. Both test statistics are asymptotically distributed as $\chi^2_1$ under the null hypothesis. If instead of a scalar $\psi$ we considered a vector of dimension $q$ ($\leq 3$), the test statistic asymptotic null distribution would be $\chi^2_q$. An alternative test is the gradient test. We shall not consider it in the Monte Carlo simulations. For details on the gradient test, see \cite{Lemonte_2016}.

\begin{table}[htp]
\centering
\caption{Null rejection rates of the LR, score and Wald tests for testing of $H_0:\alpha=0.5$ against $H_1:\alpha\neq0.5$ in the $\mathcal{BBS}(0.5, 1, 0)$ model.}
\label{T:th_penal_alfa}
\begin{small}
\begin{tabular}{c|ccc}
\hline
$n$ & LR & S & Wald  \\
\hline
\multicolumn{4}{c}{$\epsilon = 0.1$} \\
\hline
 30 & $0.2660$ & $0.2166$ & $0.4460$ \\
 50 & $0.2008$ & $0.1656$ & $0.3374$ \\
100 & $0.1472$ & $0.1288$ & $0.2392$ \\
150 & $0.1346$ & $0.1202$ & $0.2078$ \\
\hline
\multicolumn{4}{c}{$\epsilon = 0.05$} \\
\hline
 30 & $0.1632$ & $0.1116$ & $0.3850$ \\
 50 & $0.1156$ & $0.0820$ & $0.2836$ \\
100 & $0.0844$ & $0.0664$ & $0.1914$ \\
150 & $0.0716$ & $0.0574$ & $0.1524$ \\
\hline
\multicolumn{4}{c}{$\epsilon = 0.01$} \\
\hline
 30 & $0.0508$ & $0.0104$ & $0.2706$ \\
 50 & $0.0352$ & $0.0100$ & $0.1822$ \\
100 & $0.0230$ & $0.0094$ & $0.1094$ \\
150 & $0.0166$ & $0.0084$ & $0.0766$ \\
\hline
\end{tabular}
\end{small}
\end{table}

\begin{table}[htp]
\centering
\caption{Null rejection rates of the LR, score and Wald tests for testing of $H_0:\beta=1$ against $H_1:\beta\neq1$ in the $\mathcal{BBS}(0.5, 1, 0)$ model.}
\label{T:th_penal_beta}
\begin{small}
\begin{tabular}{c|cccc}
\hline
$n$ & LR & S & Wald  \\
\hline
\multicolumn{4}{c}{$\epsilon = 0.1$} \\
\hline
 30 & $0.1692$ & $0.0832$ & $0.2728$ \\
 50 & $0.1416$ & $0.0826$ & $0.2116$ \\
100 & $0.1190$ & $0.0894$ & $0.1650$ \\
150 & $0.1170$ & $0.0934$ & $0.1494$ \\
\hline
\multicolumn{4}{c}{$\epsilon = 0.05$} \\
\hline
 30 & $0.1040$ & $0.0340$ & $0.2162$ \\
 50 & $0.0802$ & $0.0370$ & $0.1480$ \\
100 & $0.0620$ & $0.0442$ & $0.1090$ \\
150 & $0.0558$ & $0.0438$ & $0.0808$ \\
\hline
\multicolumn{4}{c}{$\epsilon = 0.01$} \\
\hline
 30 & $0.0298$ & $0.0048$ & $0.1094$ \\
 50 & $0.0218$ & $0.0080$ & $0.0734$ \\
100 & $0.0112$ & $0.0076$ & $0.0354$ \\
150 & $0.0112$ & $0.0090$ & $0.0268$ \\
\hline
\end{tabular}
\end{small}
\end{table}

\begin{table}[htp]
\centering
\caption{Null rejection rates of the LR, score and Wald tests for testing $H_0:\gamma=0$ against $H_1:\gamma\neq0$ in the $\mathcal{BBS}(0.5, 1, 0)$ model.}
\label{T:th_penal_gama}
\begin{small}
\begin{tabular}{c|cccc}
\hline
$n$ & LR & S & Wald  \\
\hline
\multicolumn{4}{c}{$\epsilon = 0.1$} \\
\hline
 30 & $0.2570$ & $0.2434$ & $0.3704$ \\
 50 & $0.1974$ & $0.1920$ & $0.2896$ \\
100 & $0.1368$ & $0.1320$ & $0.2014$ \\
150 & $0.1364$ & $0.1326$ & $0.1774$ \\
\hline
\multicolumn{4}{c}{$\epsilon = 0.05$} \\
\hline
 30 & $0.1678$ & $0.1522$ & $0.2992$ \\
 50 & $0.1208$ & $0.1076$ & $0.2134$ \\
100 & $0.0860$ & $0.0762$ & $0.1454$ \\
150 & $0.0734$ & $0.0660$ & $0.1140$ \\
\hline
\multicolumn{4}{c}{$\epsilon = 0.01$} \\
\hline
 30 & $0.0520$ & $0.0334$ & $0.1632$ \\
 50 & $0.0336$ & $0.0234$ & $0.1100$ \\
100 & $0.0218$ & $0.0150$ & $0.0658$ \\
150 & $0.0154$ & $0.0114$ & $0.0474$ \\
\hline
\end{tabular}
\end{small}
\end{table}

A Monte Carlo simulation study was performed to evaluate the finite sample performances of the LR, score (denoted by S) and Wald tests in the $\mathcal{BBS}$ model. Log-likelihood maximizations were carried out using the BFGS quasi-Newton method.  The number of Monte Carlo replications is  5,000 replications, the sample sizes are $n\in\{30,50,100,150\}$ and the significance levels are $\epsilon\in\{0.1,0.05,0.01\}$. The tests were performed for each parameter of the model $\mathcal{BBS}(0.5, 1, 0)$. It is noteworthy that by testing $H_0:\gamma = 0$ against $H_1:\gamma \neq 0$ we test whether the data follows the $\mathcal{BS}$ law, i.e., the original version of the Birnbaum-Saunders distribution. The data were generated according to the model implied by the null hypothesis and samples for which convergence did not take place were discarded. 

Tables~\ref{T:th_penal_alfa} to \ref{T:th_penal_gama} contain the null rejection rates of the tests of $H_0:\alpha = 0.5$, $H_0:\beta = 1$ and $H_0:\gamma = 0$, respectively, against two-sided alternative hypotheses. We note that all tests are considerably liberal when the sample size is small (30 or 50). We also note that the score test outperforms the competition. The Wald test was the worst performer. 

The tests null rejection rates converge to the corresponding nominal levels as $n\rightarrow\infty$. Such convergence, however, is rather slow. More accurate testing inference can be achieved by using bootstrap resampling; see \cite{davison1997bootstrap}. The tests employ critical values that are estimated in the bootstrapping scheme instead of asymptotic (approximate) critical values. $B$ bootstrap samples are generated imposing the null hypothesis and the test statistic is computed for each artificial sample. The critical value of level $\epsilon\times 100\%$ is obtained as the $1-\epsilon$ upper quantile of the $B$ test statistics, i.e., of the test statistics computed using the bootstrap samples. The bootstrap tests are indicated by the subscript `pb'. We also use bootstrap resampling to estimate the Bartlett correction factor to the likelihood ratio test as proposed by \cite{rocke1989bootstrap}. The bootstrap Bartlett corrected test is indicated by the subscript `bbc'. For details on bootstrap tests, Bartlett-corrected tests and Bartlett corrections based on the bootstrap, the reader is referred to \cite{cordeiro2014introduction}. Since the Wald test proved to be considerably unreliable we shall not consider it.

\begin{table}[htp]
\centering
\caption{Null rejection rates of the bootstrap versions of the LR and score tests of $H_0:\alpha=0.5$ against $H_1:\alpha\neq0.5$ in the $\mathcal{BBS}(0.5, 1, 0)$ model.}
\label{T:RV_RVS_boot_alfa}
\begin{small}
\begin{tabular}{c|ccc}
\hline
$\epsilon$ & $\text{LR}_{pb}$ & $\text{LR}_{bbc}$ & $\text{S}_{pb}$  \\
\hline
\multicolumn{4}{c}{$n = 30$} \\
\hline
0.10 & $0.1010$ & $0.0934$ & $0.0994$ \\
0.05 & $0.0542$ & $0.0388$ & $0.0500$ \\
0.01 & $0.0100$ & $0.0038$ & $0.0082$ \\
\hline
\multicolumn{4}{c}{$n = 50$} \\
\hline
0.10 & $0.0990$ & $0.0970$ & $0.0964$ \\
0.05 & $0.0484$ & $0.0420$ & $0.0466$ \\
0.01 & $0.0096$ & $0.0058$ & $0.0108$ \\
\hline
\multicolumn{4}{c}{$n = 100$} \\
\hline
0.10 & $0.1006$ & $0.1026$ & $0.0988$ \\
0.05 & $0.0510$ & $0.0520$ & $0.0488$ \\
0.01 & $0.0106$ & $0.0102$ & $0.0106$ \\
\hline
\multicolumn{4}{c}{$n = 150$} \\
\hline
0.10 & $0.0950$ & $0.0970$ & $0.0940$ \\
0.05 & $0.0474$ & $0.0484$ & $0.0486$ \\
0.01 & $0.0096$ & $0.0096$ & $0.0110$ \\
\hline
\end{tabular}
\end{small}
\end{table}

\begin{table}[htp]
\centering
\caption{Null rejection rates of the bootstrap versions of the LR and score tests of $H_0:\beta=1$ against $H_1:\beta\neq1$ in the $\mathcal{BBS}(0.5, 1, 0)$ model.}
\label{T:RV_RVS_boot_beta}
\begin{small}
\begin{tabular}{c|ccc}
\hline
$\epsilon$ & $\text{LR}_{pb}$ & $\text{LR}_{bbc}$ & $\text{S}_{pb}$  \\
\hline
\multicolumn{4}{c}{$n = 30$} \\
\hline
0.10 & $0.1088$ & $0.1088$ & $0.1104$ \\
0.05 & $0.0496$ & $0.0490$ & $0.0514$ \\
0.01 & $0.0108$ & $0.0100$ & $0.0114$ \\
\hline
\multicolumn{4}{c}{$n = 50$} \\
\hline
0.10 & $0.1064$ & $0.1054$ & $0.1096$ \\
0.05 & $0.0532$ & $0.0528$ & $0.0538$ \\
0.01 & $0.0092$ & $0.0094$ & $0.0112$ \\
\hline
\multicolumn{4}{c}{$n = 100$} \\
\hline
0.10 & $0.1016$ & $0.1002$ & $0.1164$ \\
0.05 & $0.0538$ & $0.0524$ & $0.0574$ \\
0.01 & $0.0102$ & $0.0112$ & $0.0102$ \\
\hline
\multicolumn{4}{c}{$n = 150$} \\
\hline	
0.10 & $0.1048$ & $0.1058$ & $0.1140$ \\
0.05 & $0.0510$ & $0.0510$ & $0.0554$ \\
0.01 & $0.0098$ & $0.0106$ & $0.0104$ \\
\hline
\end{tabular}
\end{small}
\end{table}

\begin{table}[htp]
\centering
\caption{Null rejection rates of the bootstrap versions of the LR and score tests of $H_0:\gamma=0$ against $H_1:\gamma\neq0$ in the $\mathcal{BBS}(0.5, 1, 0)$ model.}
\label{T:RV_RVS_boot_gama}
\begin{small}
\begin{tabular}{c|ccc}
\hline
$\epsilon$ & $\text{LR}_{pb}$ & $\text{LR}_{bbc}$ & $\text{S}_{pb}$  \\
\hline
\multicolumn{4}{c}{$n = 30$} \\
\hline
0.10 & $0.1034$ & $0.0948$ & $0.1022$ \\
0.05 & $0.0522$ & $0.0392$ & $0.0492$ \\
0.01 & $0.0094$ & $0.0030$ & $0.0106$ \\
\hline
\multicolumn{4}{c}{$n = 50$} \\
\hline
0.10 & $0.1040$ & $0.1028$ & $0.1014$ \\
0.05 & $0.0532$ & $0.0488$ & $0.0514$ \\
0.01 & $0.0098$ & $0.0062$ & $0.0106$ \\
\hline
\multicolumn{4}{c}{$n = 100$} \\
\hline
0.10 & $0.0980$ & $0.1012$ & $0.0976$ \\
0.05 & $0.0464$ & $0.0488$ & $0.0458$ \\
0.01 & $0.0082$ & $0.0078$ & $0.0088$ \\

\hline
\multicolumn{4}{c}{$n = 150$} \\
\hline	
0.10 & $0.0986$ & $0.0992$ & $0.1014$ \\
0.05 & $0.0468$ & $0.0488$ & $0.0456$ \\
0.01 & $0.0082$ & $0.0080$ & $0.0082$ \\
\hline
\end{tabular}
\end{small}
\end{table}

\begin{figure}
\centering
\caption{Quantile-quantile plots for the LR and $\text{LR}_{bbc}$ test statistics with $n=50$, for the tests on $\alpha$ (a), on $\beta$ (b) and on $\gamma$ (c).}
\label{F:QQ_LR_LRbbc}
\includegraphics[width=\linewidth]{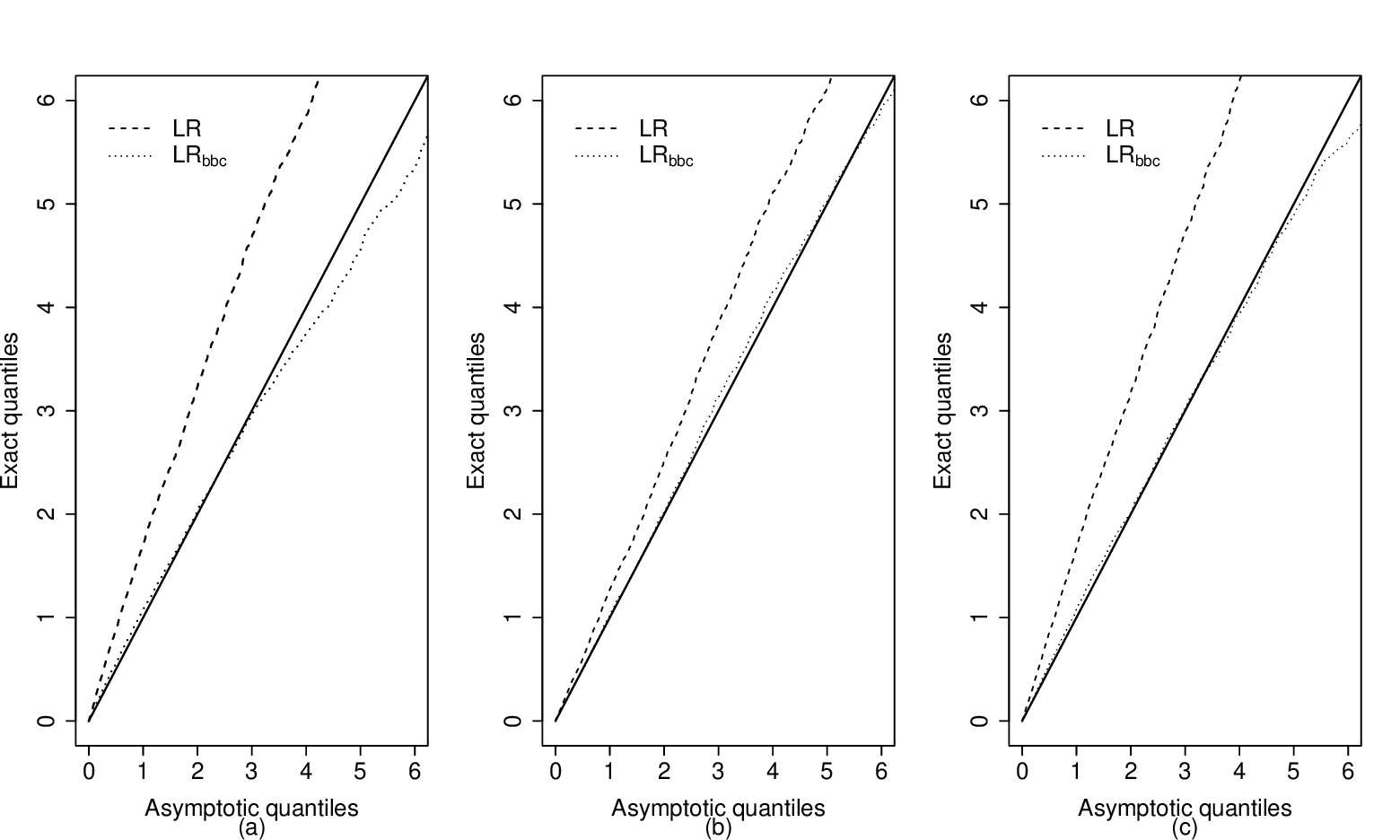}
\end{figure}

Next, we shall numerically evaluate the finite sample performances of the $\text{LR}_{pb}$, $\text{LR}_{bbc}$ and $\text{S}_{pb}$ tests under the same scenarios considered for the results presented in Tables~\ref{T:th_penal_alfa} to \ref{T:th_penal_gama}. The number of Monte Carlo replications is as before. Samples for which the optimization methods failed to reach convergence were discard, even for bootstrap samples. The same 1,000 bootstrap samples ($B=1,000$) were used in all tests. The null rejection rates of the tests used for making inferences on $\alpha$, $\beta$ and $\gamma$ are presented in Tables~\ref{T:RV_RVS_boot_alfa}, \ref{T:RV_RVS_boot_beta} and \ref{T:RV_RVS_boot_gama}, respectively. It is noteworthy that the tests size distortions are now considerably smaller. For instance, when making inference on $\alpha$ based on a sample of size $n=30$ and $\epsilon=0.05$, the LR and score null rejection rates are 16.32\% and 11.16\% (Table~\ref{T:th_penal_alfa}), whereas the corresponding figures for their bootstrap versions $\text{LR}_{\text{bp}}$, $\text{LR}_{\text{bbc}}$ and $\text{S}_{\text{bp}}$ are 5.42\%, 3.88\% and 5\%, respectively, which are much closer to 5\%. When testing restrictions on $\beta$ with $n=30$ and $\epsilon=0.10$, the LR and score null rejection rates are, respectively, 16.92\% and 8.32\% (Table~\ref{T:th_penal_beta}) whereas their bootstrap versions, $\text{LR}_{\text{bp}}$, $\text{LR}_{\text{bbc}}$ and $\text{S}_{\text{bp}}$, display null rejection rates of 10.88\%, 10.88\% and 11.04\%, respectively. Finally, when the interest lies in making inferences on $\gamma$ with $n=30$ and $\epsilon=0.01$, the LR and score null rejection rates are 5.20\% and 3.34\% (Table~\ref{T:th_penal_gama}); for the bootstrap-based tests $\text{LR}_{\text{bp}}$, $\text{LR}_{\text{bbc}}$ and $\text{S}_{\text{bp}}$ we obtain 0.94\%, 0.3\% and 1.06\%, respectively. Figure~\ref{F:QQ_LR_LRbbc} shows the quantile-quantile (QQ) plots of the LR and $\text{LR}_{bbc}$ test statistics for samples of size $n=50$. It is noteworthy that the empirical quantiles of the $\text{LR}_{bbc}$ test statistic are much more closer of the corresponding asymptotic quantiles than those of $W$. Hence, we note that testing inference in small samples can be made considerably more accurate by using bootstrap resampling.


\section{One-sided hypothesis tests}
\label{Sec:Teste_hip_unilaterais}

One-sided tests on a scalar parameter can be performed using the signed likelihood ratio (SLR) statistic, which is particularly useful in the $\mathcal{BBS}$ model since it allows practitioners to make inferences on $\gamma$ in a way that makes it possible to detect bimodality. The signed penalized likelihood ratio test statistic is 
\begin{align}
R = \sinal(\hat{\psi}-\psi_0)\sqrt{W} = \sinal(\hat{\psi}-\psi_0)\sqrt{2\{\ell^{*}(\hat{\vtheta}) - \ell^{*}(\tilde{\vtheta})\}}.
\label{E:estat_RVS}
\end{align}
The statistic $R$ is asymptotically distributed as standard normal under the null hypothesis. An advantage of the SLR test over the tests described in Section~\ref{Sec:Teste_hip_bilaterais} is that it can be used to perform two-sided and one-sided tests. In this section, we shall focus on one-sided hypothesis testing inference. Our interest lies in detecting bimodality. The null hypothesis is $H_0:\gamma \geq 0$ which is tested against $H_1: \gamma < 0$. Rejection of $H_0$ yields evidence that the data came from a bimodal distribution. On the other hand, when $H_0$ is not rejected, there is evidence that the data follow a distribution with a single mode.

Consider the sample $\vx = (x_1, \ldots, x_n)$ for a model with parameter vector $\vtheta = (\psi, \vlambda)$ of dimension $1 \times p$, where the parameter of interest $\psi$ is a scalar and the vector of nuisance parameters $\vlambda$ has dimension $1 \times (p-1)$. The test statistic $R$, given in Equation (\ref{E:estat_RVS}), is asymptotically distributed as standard normal with error of order $O(n^{-1/2})$ when the null hypothesis is true. Such an approximation may not be accurate when the sample size is small. Some analytical corrections for $R$ were proposed in the literature. They can be used to improve the test finite sample behavior.  

An important contribution was made by \cite{barndorff1986inference,barndorff1991modified}. The author proposed a correction term $\mathcal{U}$ of the form 
$$
R^* = R + \log(\mathcal{U}/R)/R,
$$ 
where $R$ represents the SLR statistic and $R^*$ is its corrected version. Let $\ell(\vtheta)$ be the log-likelihood function of the parameters. Its derivatives shall be denoted by
\begin{align*}
\ell_{\vtheta}(\vtheta) = \frac{\partial \ell(\vtheta)}{\partial \vtheta}
\quad \text{and} \quad
\ell_{\vtheta\vtheta}(\vtheta) = \frac{\partial^2\ell(\vtheta)}{\partial\vtheta\partial\vtheta^{\top}}. 
\end{align*}
The observed information matrix is given by $J_{\vtheta\vtheta}(\vtheta) = -\ell_{\vtheta\vtheta}(\vtheta)$. To obtain the correction proposed by \cite{barndorff1986inference}, the sufficient statistic has to be of the form $(\hat{\vtheta},a)$, where $\hat{\vtheta}$ is the MLE of $\vtheta$ and $a$ is an ancillary statistic. Additionally, it is necessary to compute sample space derivatives of the log-likelihood, such as 
\begin{align*}
\ell_{;\hat{\vtheta}}(\vtheta) = \frac{\partial \ell(\vtheta)}{\partial \hat{\vtheta}}
\quad \text{and} \quad
\ell_{\vtheta;\hat{\vtheta}}(\vtheta) = \frac{\partial \ell_{;\hat{\vtheta}}(\vtheta)}{\partial\vtheta^{\top}},
\end{align*}
where derivatives are taken with respect to some functions of the sample while keeping other terms fixed, as explained in \cite{severini2000likelihood}. The quantity $\mathcal{U}$ is given by
\begin{align*}
\mathcal{U} = \frac{
\left\vert\begin{array}{c}
\ell_{;\hat{\vtheta}}(\hat{\vtheta}) - \ell_{;\hat{\vtheta}}(\tilde{\vtheta}) \\
\ell_{\lambda;\hat{\vtheta}}(\tilde{\vtheta})
\end{array}\right\vert
}{\left\vert J_{\lambda\lambda}(\tilde{\vtheta}) \right\vert^{1/2}\left\vert J_{\vtheta\vtheta}(\hat{\vtheta}) \right\vert^{1/2}},
\end{align*}
where $\tilde{\vtheta}$ is the restricted MLE of $\vtheta$ and the indices indicate which components are being used in each vector or matrix.

The null distribution of $R^*$ is standard normal with error of order $O(n^{-3/2})$. Although the null distribution of $R^*$ is better approximated by  the limiting distribution than that of $R$, the computation of $\mathcal{U}$ is restricted to some specific classes of models, such as exponential family and transformation models \citep{severini2000likelihood}.

Some alternatives to $R^*$ were proposed in the literature. They approximate the sample space derivatives used in $\mathcal{U}$. For instance, approximations were obtained by \cite{diciccio1993simple}, \cite{fraser1999simple} and \cite{severini1999empirical}. They were computed by \cite{wu2004improved} for the $\mathcal{BS}$ model and by \cite{lemonte2011signed} for a Birnbaum-Saunders regression model. Other recent contributions are \cite{ferrari2014small} and \cite{smith2015penalized}. 

In this paper, we apply the approximations proposed by \cite{severini1999empirical} and \cite{fraser1999simple} for log-likelihood functions without penalization. Our interest is to evaluate the effectiveness of the corrections when applied to the statistic $R$, computed using the penalized log-likelihood function. We also compare the performances of the corrected tests to those of the SLR test and its bootstrap version.

Using the same notation as \cite{lemonte2011signed}, the approximation proposed by \cite{fraser1999simple} (denoted by $\text{SLR}_{c1}$) for $\mathcal{U}$ can be written as 
\begin{align*}
\mathcal{U}_1 = \frac{
\left\vert\begin{array}{c}
\mathbf{\Gamma}_{\vtheta} \\
\mathbf{\Psi}_{\lambda\vtheta}
\end{array}\right\vert
}{\left\vert J_{\lambda\lambda}(\tilde{\vtheta}) \right\vert^{1/2}\left\vert J_{\vtheta\vtheta}(\hat{\vtheta}) \right\vert^{1/2}},
\end{align*}
where
\begin{align*}
\mathbf{\Gamma}_{\vtheta} = [\ell_{;\vx}(\hat{\vtheta})-\ell_{;\vx}(\tilde{\vtheta})]V(\hat{\vtheta})[\ell_{\vtheta;\vx}(\hat{\vtheta})V(\hat{\vtheta})]^{-1}J_{\vtheta\vtheta}(\hat{\vtheta}),
\end{align*}
with
\begin{align*}
\mathbf{\Psi}_{\vtheta\vtheta} = \left[
\begin{array}{c}
\mathbf{\Psi}_{\psi\vtheta}\\
\mathbf{\Psi}_{\lambda\vtheta}
\end{array}\right]
= \ell_{\vtheta;\vx}(\tilde{\vtheta})V(\hat{\vtheta})[\ell_{\vtheta;\vx}(\hat{\vtheta})V(\hat{\vtheta})]^{-1}J_{\vtheta\vtheta}(\hat{\vtheta}),
\end{align*}
where $\ell_{;\vx}(\vtheta) = \partial l(\vtheta)/\partial \vx$ is a $1\times n$ vector, $\ell_{\vtheta;\vx}(\tilde{\vtheta}) = \partial^2 \ell(\vtheta)/\partial\vtheta^{\top}\partial\vx$ is a $p \times n$ matrix and
$$
V(\vtheta) = -\left[ \frac{\partial \vz(\vx;\vtheta)}{\partial\vx} \right]^{-1}\left[ \frac{\partial \vz(\vx;\vtheta)}{\partial\vtheta^{\top}} \right]
$$
is an $n \times p$ matrix, $\vz(\vx;\vtheta)$ being a vector of pivotal quantities. 

The corrected SLR statistic obtained using the approximation given by \cite{fraser1999simple} is $R_{c1} = R + \log(\mathcal{U}_{1}/R)/R$, which has asymptotic standard normal distribution with error of order $O(n^{-3/2})$ under the null hypothesis. We derived the quantities needed to obtain $\mathcal{U}_{1}$ in the $\mathcal{BBS}$ model, which are presented below. 

Consider the random variable $Y = |T|+\gamma$, where $T = \alpha^{-1}(\sqrt{X/\beta}-\sqrt{\beta/X})$ and $X\sim\mathcal{BBS}(\alpha,\beta,\gamma)$. The distribution of $Y$ is truncated standard normal with support $(\gamma, \infty)$, its distribution function being given by 
\begin{align*}
F_Y(y) = \left\{ \begin{array}{ll}
0 & \text{if } y < \gamma, \\
\frac{\Phi(y)-\Phi(\gamma)}{1-\Phi(\gamma)} & \text{if } y \geq\gamma.
\end{array}\right.
\end{align*}
Therefore, $Z = F_Y(Y)$ is uniformly distributed in the standard interval, $(0,1)$. Hence, it is a pivotal quantity that can be used for obtaining the approximations to sample space derivatives proposed by \cite{fraser1999simple}. Let $\vx = (x_1, \ldots, x_n)$ be a random $\mathcal{BBS}(\alpha, \beta, \gamma)$ sample. It follows that $\partial z_i/\partial x_j = 0$ when $i\neq j$ and $\partial z_i/\partial x_i = \phi(y_i)\, \sinal(t_i)(x_i + \beta)/[\Phi(-\gamma)2\alpha\beta^{1/2}x_i^{3/2}]$, with $t_i = \alpha^{-1}(\sqrt{x_i/\beta}-\sqrt{\beta/x_i})$. Moreover, $\partial z_i/\partial \alpha = -\phi(y_i)\, \sinal(t_i)t_i/\Phi(-\gamma)\alpha$, $\partial z_i/\partial \beta = -\phi(y_i)\, \sinal(t_i)$ $[ \sqrt{x_i/\beta} + \sqrt{\beta / x_i} ]/[\Phi(-\gamma)2\alpha\beta] $ and $\partial z_i/\partial \gamma = [\Phi(y_i) - \Phi(\gamma)]\phi(\gamma)/\Phi^2(-\gamma) +  [\phi(y_i) - \phi(\gamma)]/\Phi(-\gamma)$, where $y_i = |t_i|+\gamma$ and $z_i = F_Y(y_i)$. Therefore, $v_{\alpha i} = 2\beta^{1/2}x_i^{3/2}t_i/(x_i+\beta)$, $v_{\beta i} = x_i/\beta$ and $v_{\gamma i} = -2\alpha\beta^{1/2}x_i^{3/2}\{ [\Phi(y_i) - \Phi(\gamma)]\phi(\gamma)/\Phi(-\gamma) + \phi(y_i) - \phi(\gamma) \}/[\phi(y_i)\sinal(t_i)(x_i + \beta)]$. The vectors $\mathbf{v}_{\alpha}$, $\mathbf{v}_{\beta}$ and $\mathbf{v}_{\gamma}$ are used to form the matrix $V(\vtheta)$. For instance,  $\mathbf{v}_{\alpha} = (v_{\alpha 1}, \ldots, v_{\alpha n})^{\top}$ is a $n \times 1$ vector. Here, $V(\vtheta) = \left[ \mathbf{v}_{\alpha} \,\, \mathbf{v}_{\beta} \,\, \mathbf{v}_{\gamma} \right]$. Furthermore, we have that
\begin{align*}
\ell_{;x_i}(\vtheta) &= \frac{-3}{2x_i} + \frac{1}{x_i+\beta} - (|t_i| + \gamma)\sinal(t_i)\frac{(x_i+\beta)}{2\alpha\beta^{1/2}x_i^{3/2}}, \\
\ell_{\alpha;x_i}(\vtheta) &= \frac{\sinal(t_i)(x_i+\beta)}{2\beta^{1/2}x_i^{3/2}\alpha^2}(2|t_i|+\gamma), \\
\ell_{\beta;x_i}(\vtheta) &= \frac{(-1)}{(x_i+\beta)^2} + \frac{(x_i + \beta)}{4\alpha^2\beta^{3/2}x_i^{3/2}}\left( \frac{x_i^{1/2}}{\beta^{1/2}} + \frac{\beta^{1/2}}{x_i^{1/2}} \right) + \frac{\sinal(t_i)(|t_i|+\gamma)(x_i - \beta)}{4\alpha\beta^{3/2}x_i^{3/2}} ,\\ 
\ell_{\gamma;x_i}(\vtheta) &= -\frac{\sinal(t_i)(x_i + \beta)}{2\alpha\beta^{1/2}x_i^{3/2}}. 
\end{align*}

The method proposed by \cite{severini1999empirical} (denoted by $\text{SLR}_{c2}$) approximates the sample space derivatives by covariances of the log-likelihood function. The main idea is to use the sample to obtain the covariance values empirically. Using again the notation of \cite{lemonte2011signed}, the approximation of $\mathcal{U}$ proposed by \cite{severini1999empirical} is given by
\begin{align*}
\mathcal{U}_2 = \frac{
\left\vert\begin{array}{c}
\mathbf{\Delta}_{\vtheta} \\
\mathbf{\Sigma}_{\lambda\vtheta}
\end{array}\right\vert
}{\left\vert J_{\lambda\lambda}(\tilde{\vtheta}) \right\vert^{1/2}\left\vert J_{\vtheta\vtheta}(\hat{\vtheta}) \right\vert^{1/2}},
\end{align*}
with
\begin{align*}
\mathbf{\Delta}_{\vtheta} = [Q(\hat{\vtheta};\hat{\vtheta})-Q(\tilde{\vtheta};\hat{\vtheta})]I(\hat{\vtheta};\hat{\vtheta})^{-1}J_{\vtheta\vtheta}(\hat{\vtheta})
\end{align*}
and
\begin{align*}
\mathbf{\Sigma}_{\vtheta\vtheta} = \left[
\begin{array}{c}
\mathbf{\Sigma}_{\psi\vtheta}\\
\mathbf{\Sigma}_{\lambda\vtheta}
\end{array}\right]
= I(\tilde{\vtheta};\hat{\vtheta})I(\hat{\vtheta};\hat{\vtheta})^{-1}J_{\vtheta\vtheta}(\hat{\vtheta}),
\end{align*}
where $Q(\vtheta;\vtheta_0) = \sum_{i=1}^{n}\ell^{(i)}(\vtheta)\ell_{\vtheta}^{(i)}(\vtheta_0)^{\top}$ is an $1\times p$ vector and $I(\vtheta;\vtheta_0) = \sum_{i=1}^{n}$ $\ell_{\vtheta}^{(i)}(\vtheta)\ell_{\vtheta}^{(i)}(\vtheta_0)^{\top}$ is a $p \times p$ matrix, the index $(i)$ indicating that the quantity  corresponds to the $i$th sample observation. The corrected statistic proposed by \cite{severini1999empirical} is $R_{c2} = R + \log(\mathcal{U}_2/R)/R$. Its null distribution is standard normal with error of order $O(n^{-1})$. The score function and the observed information matrix, which can be found in \cite{olmos2015}, are used to obtain $\mathcal{U}_{2}$ in the $\mathcal{BBS}$ model.

Alternatively, bootstrap resampling can be used to obtain critical values for the SLR test. Since we test $H_0:\gamma \geq 0$ against $H_1: \gamma < 0$, the critical value of level $\epsilon\times 100\%$ is obtained as the $\epsilon$ quantile of the $B$ test statistics computed using the bootstrap samples.

\begin{table}[htp]
\centering
\caption{Null rejection rates of the SLR, $\text{SLR}_{c1}$, $\text{SLR}_{c2}$ and $\text{SLR}_{\text{bp}}$ tests of $H_0:\gamma\geq 0$ against $H_1:\gamma<0$ in a sample of size 30 of the model $\mathcal{BBS}(0.5, 1, \gamma)$.}
\label{T:th_gama_unilateral}
\begin{small}
\begin{tabular}{c|cccc}
\hline
$\epsilon$ & $\text{SLR}$ & $\text{SLR}_{c1}$ & $\text{SLR}_{c2}$ & $\text{SLR}_{bp}$   \\
\hline
\multicolumn{5}{c}{$\gamma = -1$} \\
\hline
0.10 & $0.7488$ & $0.5768$ & $0.6310$ & $0.4960$ \\
0.05 & $0.6300$ & $0.4276$ & $0.4728$ & $0.3488$ \\
0.01 & $0.3560$ & $0.1910$ & $0.2100$ & $0.1376$ \\
\hline
\multicolumn{5}{c}{$\gamma = -0.5$} \\
\hline
0.10 & $0.4634$ & $0.2766$ & $0.3300$ & $0.2248$ \\
0.05 & $0.3242$ & $0.1762$ & $0.2042$ & $0.1326$ \\
0.01 & $0.1328$ & $0.0588$ & $0.0652$ & $0.0376$ \\
\hline
\multicolumn{5}{c}{$\gamma = 0$} \\
\hline
0.10 & $0.2614$ & $0.1334$ & $0.1678$ & $0.1042$ \\
0.05 & $0.1658$ & $0.0746$ & $0.0892$ & $0.0498$ \\
0.01 & $0.0486$ & $0.0202$ & $0.0210$ & $0.0106$ \\
\hline
\multicolumn{5}{c}{$\gamma = 0.5$} \\
\hline
0.10 & $0.1546$ & $0.0722$ & $0.0928$ & $0.0526$ \\
0.05 & $0.0890$ & $0.0378$ & $0.0442$ & $0.0222$ \\
0.01 & $0.0214$ & $0.0098$ & $0.0102$ & $0.0046$ \\
\hline
\multicolumn{5}{c}{$\gamma = 1$} \\
\hline
0.10 & $0.1144$ & $0.0488$ & $0.0640$ & $0.0340$ \\
0.05 & $0.0606$ & $0.0246$ & $0.0292$ & $0.0150$ \\
0.01 & $0.0144$ & $0.0050$ & $0.0046$ & $0.0022$ \\
\hline
\end{tabular}
\end{small}
\end{table}

A simulation study was performed to evaluate the sizes and powers of the SLR, $\text{SLR}_{c1}$, $\text{SLR}_{c2}$ and $\text{SLR}_{\text{bp}}$ tests. We tested $H_0:\gamma\geq 0$ against $H_1:\gamma<0$. The true parameter values are $\gamma \in \{-1,-0.5,0,0.5,1 \}$. The most reliable tests are those with large power (i.e., higher probability of rejecting $H_0$ when $\gamma < 0$) and small size distortions. Again, 5,000 Monte Carlo replication were performed. The $\text{SLR}_{\text{bp}}$ test is based on 1,000 bootstrap samples. The simulation results are presented in Table~\ref{T:th_gama_unilateral}. The most powerful tests are SLR, $\text{SLR}_{c2}$ and $\text{SLR}_{c1}$, in that order, whereas the tests with the smallest size distortions are $\text{SLR}_{bp}$, $\text{SLR}_{c1}$ and $\text{SLR}_{c2}$. We recommend that testing inference be based on either $\text{SLR}_{c1}$ or $\text{SLR}_{c2}$, since these tests display a good balance between size and power.


\section{Nonnested hypothesis tests for the bimodal Birnbaum-Saunders model}
\label{Sec:Teste_hip_nao_encaixados} 

In the previous section we presented a test that is useful for detecting whether the data came from a bimodal $\mathcal{BBS}$ law. That was done by testing a restriction on $\gamma$. In this section we shall present tests that are useful for distinguishing between the $\mathcal{BBS}$ model and another extension of the $\mathcal{BS}$ distribution that can display bimodality. 

As noted in the Introduction, another variant of the $\mathcal{BS}$ distribution that can exhibit bimodality is the model recently discussed by \cite{owen2015revisit}, which the authors denoted by $\mathcal{GBS}_2$. Let $X\sim\mathcal{GBS}_2(\alpha, \beta, \nu)$. Its PDF is given by
\begin{align*}
g(x) = \frac{\nu}{\alpha x}\left[ \left(\frac{x}{\beta}\right)^{\nu} + \left(\frac{\beta}{x}\right)^{\nu} \right] \phi\left( \frac{1}{\alpha}\left[ \left(\frac{x}{\beta}\right)^{\nu} - \left(\frac{\beta}{x}\right)^{\nu} \right] \right), \quad x>0,
\end{align*}
where $\alpha>0$, $\beta>0$ and $\nu>0$. According to \cite{owen2015revisit}, the $\mathcal{GBS}_2$ density is bimodal when $\alpha>2$ and $\nu>2$ (simultaneously).

Therefore, when bimodality is detected the subsequent data analysis may be carried out with either the $\mathcal{BBS}$ distribution or the $\mathcal{GBS}_2$ model. It would then be useful to have a hypothesis test that could be used to distinguish between the two models. Obviously, the tests discussed so far cannot be used to that end. $\mathcal{BS}$ model selection criteria were considered by \cite{leiva2015birnbaum} and \cite{leiva2015modeling}. Model selection is usually based on the Bayes factor and also on the Schwarz and Akaike information criteria. We shall use a different approach: we shall develop tests for nonnested hypotheses. Notice that the $\mathcal{GBS}_2$ distribution cannot be obtained from the $\mathcal{BBS}$ distribution by imposing restrictions on the model parameters, and vice-versa. Hence, the two models are not nested. 

The literature of nonnested models began with \cite{cox1961tests,cox1962further}. The author introduced likelihood ratio tests for some nonnested models. His main results were generalized by \cite{vuong1989likelihood}, who considered nested, nonnested and overlapping models and derived the required asymptotics. For nonnested models, \cite{vuong1989likelihood} established the relationship between the likelihood ratio statistic and the Kullback-Leibler information. Let $F$ and $G$ be competing nonnested models. The author presented a test of the null hypothesis $H_0$ that both models are equivalent, the alternative hypotheses being: $H_f$: model $F$ is better and $H_g$: model $G$ is better. An alternative approach for testing nonnested models was considered by \cite{williams1970discrimination} and \cite{lewis2011unified}. The authors only considered tests of the hypothesis $H_f$ and $H_g$. They proposed to consider $H_f$ and $H_g$ sequentially. 

We shall consider the hypothesis involving the $\mathcal{BBS}$ and $\mathcal{GBS}_2$ models as: 
\begin{itemize}
\item $H_f$ - the data came from the $\mathcal{BBS}$ distribution,
\item $H_g$ - the data came from the $\mathcal{GBS}_2$ distribution.
\end{itemize}
The test statistic we consider is the following likelihood ratio statistic: 
$$
W_{ne} = \log\left( \frac{\hat{f}}{\hat{g}} \right) = \hat{\ell}_f - \hat{\ell}_g,
$$
where $\hat{f}$ and $\hat{g}$ denote the likelihood functions of the $\mathcal{BBS}$ and $\mathcal{GBS}_2$ models, respectively, evaluated at the respective maximum likelihood estimates, $\ell$ representing the log-likelihood function of the model indicated by its index. Then, for a given sample $\vx$, a large positive value of $W_{ne}$ yields evidence in favor $H_f$ and against $H_g$; on the other hand, a large negative value of $W_{ne}$ favors $H_g$. The $\mathcal{BBS}$ parameters are estimated using the penalized log-likelihood function and those of $\mathcal{GBS}_2$ are estimated using the standard log-likelihood function.

In the test introduced by \cite{vuong1989likelihood} for nonnested models, the test statistic asymptotic null distribution is standard normal. In some Monte Carlo simulations not reported here, this test, based on asymptotic critical values, indicated the models equivalence too frequently. Since the test is based on a large sample approximation, superior finite sample performance can be achieved by using bootstrap resampling. Application of the bootstrap method is not, however, straightforward for the test at hand because one would need to define an model equivalent to $\mathcal{BBS}$ and $\mathcal{GBS}_2$ in order to generate pseudo-samples under the null hypothesis. Thus, an approach similar to the one employed by \cite{lewis2011unified}, which only considers the hypotheses $H_f$ and $H_g$ in the test, will be used in this paper. The null hypothesis $H_f$ can be tested, using bootstrap resampling, as follows: 
\begin{enumerate}
\item Compute $W_{ne}$ using sample $\vx$;
\item With the MLEs of the parameters from the $\mathcal{BBS}$ model, generate a bootstrap sample $\vx^{*}$, and then compute $W_{ne}^*$ using that sample;
\item Execute step 2 $B$ times and obtain the bootstrap $p$-value: $p_b = \frac{\#\{W_{ne}^*<W_{ne}\} + 1}{B+1}$.
\end{enumerate}
Hence, at the $\epsilon\times100\%$ significance level, $H_f$ is rejected if $p_b < \epsilon$, i.e., we reject the hypothesis that the data originated from the $\mathcal{BBS}$ law and conclude that the $\mathcal{GBS}_2$ distribution is more adequate. Similar testing inference can be performed by taking $H_g$ as the null hypothesis. The test is carried out as follows: 
\begin{enumerate}
\item Compute $W_{ne}$ using sample $\vx$;
\item With the MLEs of the parameters from the $\mathcal{GBS}_2$ model, generate a bootstrap sample $\vx^{*}$, and then compute $W_{ne}^*$ using that sample;
\item Execute step 2 $B$ times and obtain the bootstrap $p$-value: $p_b = \frac{\#\{W_{ne}^*>W_{ne}\} + 1}{B+1}$.
\end{enumerate}
It is noteworthy that step 3 is different from the corresponding step in the first procedure, since the rejection region changes when we consider $H_g$ as the null hypothesis. Again, at the $\epsilon\times100$\% significance level, the null hypothesis is rejected if $p_b < \epsilon$, but now that means that we reject the hypothesis $H_g$ according to which the data came from the $\mathcal{GBS}_2$ distribution and conclude that the $\mathcal{BBS}$ model is more adequate.

The problem with this approach is that four inference results can happen, as noted by \cite{williams1970discrimination} and \cite{lewis2011unified}:
\begin{itemize}
\item[$R1$] The two null hypotheses, $H_f$ and $H_g$, are not rejected, and we conclude that both models are adequate;
\item[$R2$] We do not reject $H_f$, but $H_g$ is rejected, thus indicating that the $\mathcal{BBS}$ model is more adequate;
\item[$R3$] We do not reject $H_g$, but $H_f$ is rejected, thus indicating that the $\mathcal{GBS}_2$ model is more adequate;
\item[$R4$] We reject both null hypotheses, $H_f$ and $H_g$, and conclude that neither model is adequate;
\end{itemize}

Under some regularity conditions, \cite{vuong1989likelihood} has shown that, in nonnested models, an adjusted likelihood ratio test statistic tends to infinity under $H_f$ when $n\rightarrow \infty$ and that under $H_g$ it tends to minus infinity when $n\rightarrow \infty$. That way, the test statistic tends to indicate the correct model as the sample size increases. Therefore, when result $R1$ is reached, model selection can be based on $W_{ne}$: the $\mathcal{BBS}$ distribution is selected if $W_{ne}>0$ and the $\mathcal{GBS}_2$ distribution is selected if $W_{ne}<0$.

\begin{figure}
\centering
\caption{Densities $\mathcal{BBS}(0.2,1,-1)$ (solid line) and $\mathcal{GBS}_2(5, 1, 5)$ (dashed line).}
\label{F:fdp_bbs_gbs2}
\includegraphics[scale=.30]{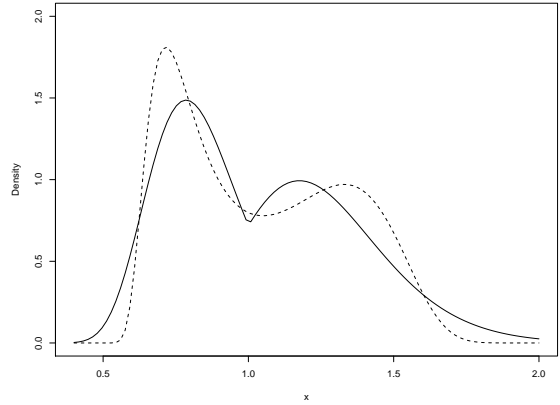}
\end{figure}

\begin{table}[htp]
\centering
\caption{Proportions of outcomes of the test of $H_f$ against $H_g$ when the data generating function is $\mathcal{BBS}(0.2, 1, -1)$ (first four columns), and proportions of $\mathcal{BBS}$ and $\mathcal{GBS}_2$ model selection and porportion of no model selected. The test significance level is $\epsilon$.}
\label{T:th_bbs_gbs2_Hf}
\begin{small}
\begin{tabular}{c|cccc|ccc}
\hline
$\epsilon$ & $R1$ & $R2$ & $R3$ & $R4$ & $\mathcal{BBS}$ & $\mathcal{GBS}_2$ & None \\
\hline
\multicolumn{8}{c}{$n = 30$} \\
\hline
0.10 & $0.6676$ & $0.2378$ & $0.0946$ & $0.0000$ & $0.4354$ & $0.5646$ & $0.0000$ \\
0.05 & $0.8284$ & $0.1322$ & $0.0394$ & $0.0000$ & $0.4354$ & $0.5646$ & $0.0000$ \\
0.01 & $0.9650$ & $0.0292$ & $0.0058$ & $0.0000$ & $0.4354$ & $0.5646$ & $0.0000$ \\
\hline
\multicolumn{8}{c}{$n = 50$} \\
\hline
0.10 & $0.5454$ & $0.3578$ & $0.0968$ & $0.0000$ & $0.5574$ & $0.4426$ & $0.0000$ \\
0.05 & $0.7230$ & $0.2326$ & $0.0444$ & $0.0000$ & $0.5572$ & $0.4428$ & $0.0000$ \\
0.01 & $0.9174$ & $0.0754$ & $0.0072$ & $0.0000$ & $0.5572$ & $0.4428$ & $0.0000$ \\
\hline
\multicolumn{8}{c}{$n = 100$} \\
\hline
0.10 & $0.3420$ & $0.5538$ & $0.1036$ & $0.0006$ & $0.6952$ & $0.3042$ & $0.0006$ \\
0.05 & $0.5344$ & $0.4138$ & $0.0518$ & $0.0000$ & $0.6940$ & $0.3060$ & $0.0000$ \\
0.01 & $0.8024$ & $0.1884$ & $0.0092$ & $0.0000$ & $0.6940$ & $0.3060$ & $0.0000$ \\
\hline
\multicolumn{8}{c}{$n = 150$} \\
\hline
0.10 & $0.1920$ & $0.6908$ & $0.1138$ & $0.0034$ & $0.7682$ & $0.2284$ & $0.0034$ \\
0.05 & $0.3772$ & $0.5646$ & $0.0582$ & $0.0000$ & $0.7656$ & $0.2344$ & $0.0000$ \\
0.01 & $0.6720$ & $0.3118$ & $0.0162$ & $0.0000$ & $0.7654$ & $0.2346$ & $0.0000$ \\
\hline
\end{tabular}
\end{small}
\end{table}

\begin{table}[htp]
\centering
\caption{Proportions of outcomes of the test of $H_f$ against $H_g$ when the data generating function is $\mathcal{GBS}_2(5, 1, 5)$ (first four columns), and proportions of $\mathcal{BBS}$ and $\mathcal{GBS}_2$ model selection and proportion of no model selected. The test significance level is $\epsilon$.}
\label{T:th_bbs_gbs2_Hg}
\begin{small}
\begin{tabular}{c|cccc|ccc}
\hline
$\epsilon$ & $R1$ & $R2$ & $R3$ & $R4$ & $\mathcal{BBS}$ & $\mathcal{GBS}_2$ & None \\
\hline
\multicolumn{8}{c}{$n = 30$} \\
\hline
0.10 & $0.4438$ & $0.1080$ & $0.4482$ & $0.0000$ & $0.1784$ & $0.8216$ & $0.0000$ \\
0.05 & $0.6658$ & $0.0538$ & $0.2804$ & $0.0000$ & $0.1784$ & $0.8216$ & $0.0000$ \\
0.01 & $0.9040$ & $0.0112$ & $0.0848$ & $0.0000$ & $0.1784$ & $0.8216$ & $0.0000$ \\
\hline
\multicolumn{8}{c}{$n = 50$} \\
\hline
0.10 & $0.2204$ & $0.1010$ & $0.6786$ & $0.0000$ & $0.1270$ & $0.8730$ & $0.0000$ \\
0.05 & $0.4502$ & $0.0466$ & $0.5032$ & $0.0000$ & $0.1238$ & $0.8762$ & $0.0000$ \\
0.01 & $0.7624$ & $0.0090$ & $0.2286$ & $0.0000$ & $0.1238$ & $0.8762$ & $0.0000$ \\
\hline
\multicolumn{8}{c}{$n = 100$} \\
\hline
0.10 & $0.0146$ & $0.0662$ & $0.8804$ & $0.0388$ & $0.0672$ & $0.8940$ & $0.0388$ \\
0.05 & $0.1068$ & $0.0560$ & $0.8350$ & $0.0022$ & $0.0712$ & $0.9266$ & $0.0022$ \\
0.01 & $0.4000$ & $0.0112$ & $0.5888$ & $0.0000$ & $0.0674$ & $0.9326$ & $0.0000$ \\
\hline
\multicolumn{8}{c}{$n = 150$} \\
\hline
0.10 & $0.0002$ & $0.0158$ & $0.8968$ & $0.0872$ & $0.0158$ & $0.8970$ & $0.0872$ \\
0.05 & $0.0136$ & $0.0310$ & $0.9344$ & $0.0210$ & $0.0320$ & $0.9470$ & $0.0210$ \\
0.01 & $0.1610$ & $0.0122$ & $0.8268$ & $0.0000$ & $0.0322$ & $0.9678$ & $0.0000$ \\
\hline
\end{tabular}
\end{small}
\end{table}

A simulation study was performed to evaluate the performances of the nonnested hypothesis tests involving the $\mathcal{BBS}$ and $\mathcal{GBS}_2$ distributions. The models considered were $\mathcal{BBS}(0.2,1,-1)$ and $\mathcal{GBS}_2(5,1,5)$. Figure~\ref{F:fdp_bbs_gbs2} shows the two densities. The number of Monte Carlo replications used was 5,000. First, we considered the case in which the true distribution is $\mathcal{BBS}(0.2,1,-1)$; in each replication, $B=1,000$ bootstrap samples were generated under $H_f$ and other $B=1,000$ bootstrap samples were generated under $H_g$, thus reaching one of the previously indicated results ($R1$, $R2$, $R3$ or $R4$) in each replication. Table~\ref{T:th_bbs_gbs2_Hf} contains the proportions of times that each inference was reached and also the proportions of times each distribution was chosen as the most suitable model, which is: the $\mathcal{BBS}$ model when we obtain result $R_1$ and $W_{ne}>0$ or when we obtain result $R_2$; the $\mathcal{GBS}_2$ model when we obtain result $R_1$ and $W_{ne}<0$ or when we obtain result $R_3$; none of the considered distributions under result $R_4$. The same procedure was used when the true model was the $\mathcal{GBS}_2(5,1,5)$ distribution, the corresponding results being presented in Table~\ref{T:th_bbs_gbs2_Hg}.

Based on the figures in Table~\ref{T:th_bbs_gbs2_Hf}, we note that the null rejection rates of the true hypothesis ($H_f$) are close to the nominal levels. For instance, when $n=30$ and $\epsilon=0.10$, by adding the cells corresponding to $R3$ and $R4$, we see that the rejection rate of $H_f$ is 9.46\%, which is close to the test nominal level. We also note that for small or moderate sample sizes the tests tend to indicate equivalence of both models, but as $n$ increases, the tests tend to indicate the $\mathcal{BBS}$ model as the most suitable model with increasing frequency. For example, when $\epsilon=0.05$, in the column that corresponds to $R2$, that happens for 13.22\% of the samples when $n=30$, whereas for $n=150$, for 56.46\% of the samples the $\mathcal{BBS}$ distribution is considered the most adequate model. This can also be observed in the fifth column, where we can see that as $n$ increases, the $\mathcal{BBS}$ model is selected more frequently.

Table~\ref{T:th_bbs_gbs2_Hg} contains the results obtained by taking $H_g$ as the true null hypothesis. Once again, the null rejection rates stayed close of the nominal levels. When $n=30$ and $\epsilon=0.05$, the sum of the cells relative to $R2$ and $R4$ equals 5.38\%, which is quite close to the test significance level. Moreover, we note that the results in Table~\ref{T:th_bbs_gbs2_Hg} are superior to those obtained under $H_f$. In the column corresponding to $R3$ we find the proportions of samples for which the $\mathcal{GBS}_2$ law was chosen as the most adequate model by disregarding the sign of $W_{ne}$. When $\epsilon=0.05$, for $n=30$ that happens for 28.04\% of all samples and for $n=150$ that happens for 93.44\% of the samples, a performance which is superior to that observed when $H_f$ was the true model. This can also be seen in the sixth column of the table. The proportions of samples for which the $\mathcal{GBS}_2$ distribution is correctly chosen are higher than the corresponding values in the fifth column of Table~\ref{T:th_bbs_gbs2_Hf}.

Therefore, we conclude that the bootstrap-based nonnested hypothesis tests used for distinguishing between the $\mathcal{BBS}$ and $\mathcal{GBS}_2$ models perform well. For both distributions the null rejection rates are close to the nominal levels. We also note that, as $n$ increases, the tests tend to single out the correct model with increasing frequency. We also note that the tests perform better when the true model is the $\mathcal{GBS}_2$ distribution.


\section{Empirical applications}
\label{Sec:aplicacao}

\subsection{Runoff amounts}

We shall now return to the data briefly described in Section~\ref{Se:Introducao}, which we used to illustrate the problem of nonconvergence of optimization processes during $\mathcal{BBS}$ parameter estimation. The data, provided by \cite{folks1978inverse}, consist of 25 runoff amounts at Jug Bridge, in Maryland. Table~\ref{T:runoff_descritiva} contains some descriptive statistics. We note that the data have  large kurtosis, i.e., they come from a leptokurtic distribution, and have small variance, which might be indicative that the data are concentrated around the mean and median values. These characteristics may suggest that the data came from a unimodal population. 

\begin{table}[htp]
\centering
\caption{Descriptive statistics for the runoff data.}
\label{T:runoff_descritiva}
\begin{tabular}{ccccccc}
\hline
 min & max & median & mean & variance  & asymmetry & kurtosis \\
 \hline
0.17 & 2.92 & 0.7 & 0.84 & 0.3459 & 1.7953 & 6.7493 \\
\hline
\end{tabular}
\end{table}

The models fitted to the data were $\mathcal{BS}(\alpha, \beta)$ and $\mathcal{BBS}(\alpha, \beta, \gamma)$. For the former, we obtained $\hat{\alpha} = 0.66$ $(0.0936)$ and $\hat{\beta} = 0.69$ $(0.0865)$; the numbers in parentheses are standard errors. For the second model, the maximum likelihood estimates could not be obtained because the optimization algorithm failed to converge. As shown in Figure~\ref{F:runoff_logliks}a, the log-likelihood function has a region which is apparently flat for some values of the parameters $\alpha$ and $\gamma$, with the value of $\beta$ being fixed at $0.69$. In contrast, the $\mathcal{BBS}$ penalized maximum likelihood estimates ($\text{MLE}_{\text{p}}$) were easily obtained: $\hat{\alpha} = 0.63$ $(0.2287)$, $\hat{\beta} = 0.69$ $(0.0817)$ and $\hat{\gamma} = -0.13$ $(0.8449)$. Notice that the standard error of $\hat{\gamma}$ is large relative to the point estimate which indicates that the $\mathcal{BS}$ model is adequate.

\begin{figure}
\centering
\caption{Histogram of the runoff data with the fitted densities obtained with $\mathcal{BS}(0.66,0.69)$ (dashed line) and $\mathcal{BBS}(0.63,0.69,-0.13)$ (dotted line).}
\label{F:runoff_fits}
\includegraphics[scale=.50]{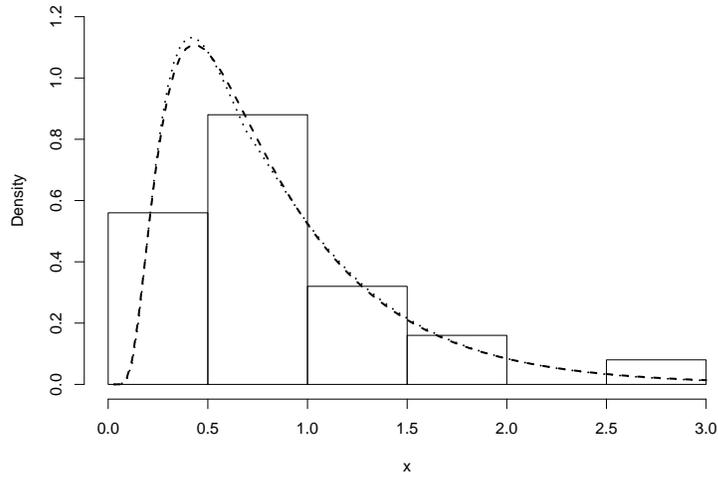}
\end{figure}

Figure~\ref{F:runoff_fits} contains the data histogram and the two fitted densities. It is noteworthy that the fitted densities are very similar. Since the $\mathcal{BS}$ distribution is simpler than the $\mathcal{BBS}$ distribution, it is to preferred. As a confirmation check, we tested the null hypothesis $H_0:\gamma=0$ against a two-sided alternative. The $p$-values of the LR, score, Wald, $\text{LR}_{pb}$, $\text{LR}_{bbc}$ and $\text{S}_{pb}$ tests were equal to 0.85, 0.81, 0.87, 0.92, 0.89 and 0.91, respectively. Therefore, there is strong evidence that the $\mathcal{BS}$ fit is adequate.

\subsection{Depressive condition data}

The second empirical application uses data on the emotional condition of 134 children. In particular, the interest lies in modeling depression measures. The data were analyzed, for example, by \cite{leiva2010unified} and \cite{balakrishnan2011some}. In both papers, mixtures of distributions were used.  

\begin{figure}
\centering
\caption{Histogram of the depressive condition data with the fitted densities obtained with $\mathcal{BS}(0.60,7.58)$ (solid line), $\mathcal{BBS}(0.42, 7.54, -0.85)$ (dashed line) and $\mathcal{GBS}_2(2.38, 7.74, 1.53)$ (dotted line).}
\label{F:depressive_fits}
\includegraphics[scale=.50]{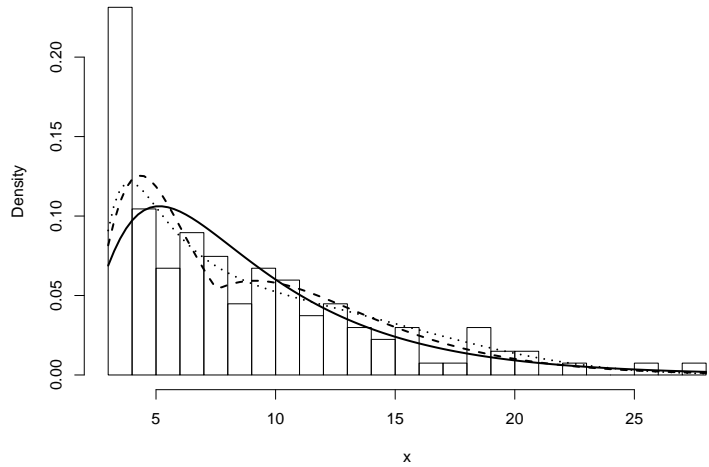}
\end{figure}

Table~\ref{T:depressive_descritiva} presents some descriptive statistics. The data are right-skewed, leptokurtic and highly dispersed. The following models were fitted: $\mathcal{BS}$, $\mathcal{BBS}$ and $\mathcal{GBS}_2$. For each model, we computed the Akaike (AIC) and Schwarz (BIC) information criteria. The $\mathcal{BS}$ estimates (standard errors in parentheses) are $\hat{\alpha}=0.603$ $(0.0368)$ and $\hat{\beta}=7.58$ $(0.3773)$, with AIC and BIC values of 780.09 and 785.89, respectively. For the $\mathcal{BBS}$ model, $\hat{\alpha}=0.42$ $(0.0481)$, $\hat{\beta}=7.54$ $(0.2645)$ and $\hat{\gamma}=-0.85$ $(0.2569)$, with AIC and BIC values of 776.26 and 784.95, respectively. For the $\mathcal{GBS}_2$ model the point estimates are $\hat{\alpha}=2.38$ $(0.6290)$, $\hat{\beta}=7.74$ $(0.3045)$ and $\hat{\nu}=1.53$ $(0.2525)$, with AIC and BIC values of 771.78 and 780.47, respectively. The data histogram and the fitted densities are presented in Figure~\ref{F:depressive_fits}.

The information criteria favor the $\mathcal{GBS}_2$ model, followed by the $\mathcal{BBS}$ law. We performed a nonnested hypothesis test to distinguish between the two models. For these data, $W_{ne}=-0.0167$ with $p$-value of 0.0189 under $H_f$ ($\mathcal{BBS}$ is taken to be the true model) and $p$-value of 0.6453 under $H_g$ ($\mathcal{GBS}_2$ is assumed to be the true model). Hence, we select $\mathcal{GBS}_2$ model. 

\begin{table}[htp]
\centering
\caption{Descriptive statistics for the depressive condition data.}
\label{T:depressive_descritiva}
\begin{tabular}{ccccccc}
\hline
 min & max & median & mean & variance  & asymmetry & kurtosis \\
 \hline
3 & 28 & 8 & 8.96 & 28.73 & 1.11 & 3.88 \\
\hline
\end{tabular}
\end{table}

\subsection{Adhesive strength}

The third data set analyzed is provided by \cite{ehsani1996design} and was also analyzed by \cite{olmos2015}, who used the $\mathcal{BBS}$ distribution in the analysis. The data consist of 48 observations on the adhesive strength to concrete of bars reinforced with glass fiber. Some descriptive statistics are presented in Table~\ref{T:adhesive_descritiva}. Notice the large  kurtosis coefficient (in excess of 5), the positive asymmetry and also the fact that the variance is considerably larger than both the mean and the median.

\begin{table}[htp]
\centering
\caption{Descriptive statistics for the adhesive strength data.}
\label{T:adhesive_descritiva}
\begin{tabular}{ccccccc}
\hline
 min & max & median & mean & variance  & asymmetry & kurtosis \\
 \hline
3.4 & 25.5 & 5.95 & 8.08 & 23.7017 & 1.448 & 5.0345 \\
\hline
\end{tabular}
\end{table}

\begin{figure}
\centering
\caption{Histogram of the adhesive strength data with the fitted densities obtained with $\mathcal{BS}(0.54,7.05)$ (solid line), $\mathcal{BBS}(0.31, 7.39, -1.38)$ (dashed line) and $\mathcal{GBS}_2(3.19, 8.05, 1.99)$ (dotted line).}
\label{F:adhesive_fits}
\includegraphics[scale=.50]{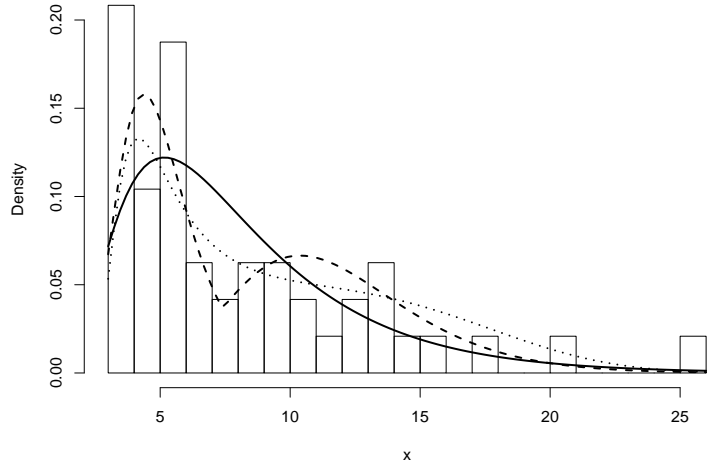}
\end{figure}

Once again, the $\mathcal{BS}$, $\mathcal{BBS}$ and $\mathcal{GBS}_2$ models were fitted to the data. The $\mathcal{BS}$ point estimates are $\hat{\alpha}=0.54$ $(0.0553)$ and $\hat{\beta}=7.05$ $(0.5316)$, the AIC and BIC values being 264.52 and 268.26, respectively. For the $\mathcal{BBS}$ model, $\hat{\alpha}=0.31$ $(0.0460)$, $\hat{\beta}=7.39$ $(0.3162)$ and $\hat{\gamma}=-1.38$ $(0.3525)$, with AIC and BIC values of 260.06 and 265.67, respectively. Finally, for the $\mathcal{GBS}_2$ distribution, $\hat{\alpha}=3.19$ $(1.5536)$, $\hat{\beta}=8.05$ $(0.5371)$ and $\hat{\nu}=1.99$ $(0.5203)$, with AIC and BIC values of 262.26 and 267.88, respectively. The data histogram and the fitted densities are shown in Figure~\ref{F:adhesive_fits}.

For this data set, the best fit according to the information criteria is the $\mathcal{BBS}$ fit, followed by  $\mathcal{GBS}_2$. The nonnested hypothesis test statistic is $W_{ne}=0.0229$, with $p$-value of 0.6543 under $H_f$ and with $p$-value of 0.0489 under $H_g$. Thus, there is substantial evidence that the $\mathcal{BBS}$ distribution is the most adequate model for these data.

For the $\mathcal{BBS}$ model, we tested unimodality versus bimodality. The hypotheses in the test were $H_0:\gamma\geq 0$ and $H_1:\gamma<0$. The $p$-values of the SLR, $\text{SLR}_{c1}$, $\text{SLR}_{c2}$ and $\text{SLR}_{\text{bp}}$ tests were 0.0002, 0.0007, 0.0006 and 0.002, respectively. Hence, all tests reject $H_0$ in favor of $H_1$, which implies that there is strong evidence in favor of $\gamma<0$, i.e., that the data came from a bimodal $\mathcal{BBS}$ law.


\section{Concluding remarks}
\label{Sec:Conclusao}

Optimization processes may often fail to reach convergence when used to obtain maximum likelihood estimates of the parameters that index the $\mathcal{BBS}$ model, an interesting extension of the well known Birnbaum-Saunders model that may display bimodality. A penalization of the log-likelihood function that uses the Jeffreys prior was proposed. Alternative strategies to circumvent the problem were also explored. Overall, the best results were obtained using a penalized log-likelihood function based on a modified version of the Jeffreys prior. 

We also considered hypothesis testing inference. For that, we used the $\mathcal{BBS}$ log-likelihood function penalized using the modified Jeffreys prior. The likelihood ratio, score and Wald tests were shown to be liberal in small samples, the Wald test being the worst performer. We have also shown that their bootstrap variants are typically quite accurate. One-sided tests based on the signed likelihood ratio statistic were also considered. We derived analytical corrections to the test statistic and also used bootstrap resampling. Overall, the analytically corrected tests displayed superior performance. We also developed tests for distinguishing between nonnested models. Our interested lied in distinguishing between the $\mathcal{BBS}$ model and an alternative version of the Birnbaum-Saunders distribution that also exhibits bimodality. Since in this case there are two distributions, the test was performed by considering two null hypotheses. It was shown that bootstrap-based nonnested testing inference can be quite accurate. 

Three empirical applications were presented and discussed. In the first application, it was not possible to obtain the $\mathcal{BBS}$ maximum likelihood point estimates since it was not possible to maximize the log-likelihood function. Parameter estimates were easily obtained when the penalized log-likelihood function proposed in this paper was used.  
Other two applications were presented. In one of them, the $\mathcal{BBS}$ model was selected as the best model and it was shown that there was substantial evidence that the true data generating process is bimodal.


\section*{Acknowledgements}

We gratefully acknowledge partial financial support from CAPES and CNPq. We also thank two anonymous referees for comments and suggestions.

\section*{References}

\bibliography{bibfile}

\begin{thebibliography}{53}
\expandafter\ifx\csname natexlab\endcsname\relax\def\natexlab#1{#1}\fi
\providecommand{\bibinfo}[2]{#2}
\ifx\xfnm\relax \def\xfnm[#1]{\unskip,\space#1}\fi
\bibitem[{Azzalini and Arellano-Valle(2013)}]{azzalini2013maximum}
\bibinfo{author}{A.~Azzalini}, \bibinfo{author}{R.B. Arellano-Valle},
  \bibinfo{title}{Maximum penalized likelihood estimation for skew-normal and
  skew-t distributions}, \bibinfo{journal}{J. Stat. Plan. Infer.}
  \bibinfo{volume}{143} (\bibinfo{year}{2013}) \bibinfo{pages}{419--433}.
\bibitem[{Balakrishnan et~al.(2011)Balakrishnan, Gupta, Kundu, Leiva and
  Sanhueza}]{balakrishnan2011some}
\bibinfo{author}{N.~Balakrishnan}, \bibinfo{author}{R.C. Gupta},
  \bibinfo{author}{D.~Kundu}, \bibinfo{author}{V.~Leiva},
  \bibinfo{author}{A.~Sanhueza}, \bibinfo{title}{On some mixture models based
  on the {B}irnbaum--{S}aunders distribution and associated inference},
  \bibinfo{journal}{J. Stat. Plann. Infer.} \bibinfo{volume}{141}
  (\bibinfo{year}{2011}) \bibinfo{pages}{2175--2190}.
\bibitem[{Balakrishnan et~al.(2009)Balakrishnan, Leiva, Sanhueza and
  Vilca}]{balakrishnan2009estimation}
\bibinfo{author}{N.~Balakrishnan}, \bibinfo{author}{V.~Leiva},
  \bibinfo{author}{A.~Sanhueza}, \bibinfo{author}{F.~Vilca},
  \bibinfo{title}{Estimation in the {B}irnbaum--{S}aunders distribution based
  on scale-mixture of normals and the {EM}-algorithm}, \bibinfo{journal}{Sort}
  \bibinfo{volume}{33} (\bibinfo{year}{2009}) \bibinfo{pages}{171--192}.
\bibitem[{Barndorff-Nielsen(1986)}]{barndorff1986inference}
\bibinfo{author}{O.E. Barndorff-Nielsen}, \bibinfo{title}{Inference on full or
  partial parameters based on the standardized signed log likelihood ratio},
  \bibinfo{journal}{Biometrika} \bibinfo{volume}{73} (\bibinfo{year}{1986})
  \bibinfo{pages}{307--322}.
\bibitem[{Barndorff-Nielsen(1991)}]{barndorff1991modified}
\bibinfo{author}{O.E. Barndorff-Nielsen}, \bibinfo{title}{Modified signed log
  likelihood ratio}, \bibinfo{journal}{Biometrika} \bibinfo{volume}{78}
  (\bibinfo{year}{1991}) \bibinfo{pages}{557--563}.
\bibitem[{Birnbaum and Saunders(1969{\natexlab{a}})}]{birnbaum1969b}
\bibinfo{author}{Z.W. Birnbaum}, \bibinfo{author}{S.C. Saunders},
  \bibinfo{title}{Estimation for a family of life distributions with
  applications to fatigue}, \bibinfo{journal}{J. Appl. Probab.}
  \bibinfo{volume}{6} (\bibinfo{year}{1969}{\natexlab{a}})
  \bibinfo{pages}{328--347}.
\bibitem[{Birnbaum and Saunders(1969{\natexlab{b}})}]{birnbaum1969a}
\bibinfo{author}{Z.W. Birnbaum}, \bibinfo{author}{S.C. Saunders},
  \bibinfo{title}{A new family of life distributions}, \bibinfo{journal}{J.
  Appl. Probab.} \bibinfo{volume}{6} (\bibinfo{year}{1969}{\natexlab{b}})
  \bibinfo{pages}{319--327}.
\bibitem[{Bourguignon et~al.(2014)Bourguignon, Silva and
  Cordeiro}]{bourguignon2014new}
\bibinfo{author}{M.~Bourguignon}, \bibinfo{author}{R.B. Silva},
  \bibinfo{author}{G.M. Cordeiro}, \bibinfo{title}{A new class of fatigue life
  distributions}, \bibinfo{journal}{J. Stat. Comput. Simul.}
  \bibinfo{volume}{84} (\bibinfo{year}{2014}) \bibinfo{pages}{2619--2635}.
\bibitem[{Cordeiro and Cribari-Neto(2014)}]{cordeiro2014introduction}
\bibinfo{author}{G.M. Cordeiro}, \bibinfo{author}{F.~Cribari-Neto},
  \bibinfo{title}{An Introduction to Bartlett Correction and Bias Reduction},
  \bibinfo{publisher}{Springer}, \bibinfo{address}{New York},
  \bibinfo{year}{2014}.
\bibitem[{Cordeiro and Lemonte(2014)}]{cordeiro2014exponentiated}
\bibinfo{author}{G.M. Cordeiro}, \bibinfo{author}{A.J. Lemonte},
  \bibinfo{title}{The exponentiated generalized {B}irnbaum--{S}aunders
  distribution}, \bibinfo{journal}{Appl. Math. Comput.} \bibinfo{volume}{247}
  (\bibinfo{year}{2014}) \bibinfo{pages}{762--779}.
\bibitem[{Cordeiro et~al.(2013)Cordeiro, Lemonte and
  Ortega}]{cordeiro2013extended}
\bibinfo{author}{G.M. Cordeiro}, \bibinfo{author}{A.J. Lemonte},
  \bibinfo{author}{E.M. Ortega}, \bibinfo{title}{An extended fatigue life
  distribution}, \bibinfo{journal}{Statistics} \bibinfo{volume}{47}
  (\bibinfo{year}{2013}) \bibinfo{pages}{626--653}.
\bibitem[{Cox(1961)}]{cox1961tests}
\bibinfo{author}{D.R. Cox}, \bibinfo{title}{Tests of separate families of
  hypotheses}, in: \bibinfo{booktitle}{Proceedings of the Fourth Berkeley
  Symposium on Mathematical Statistics and Probability},
  volume~\bibinfo{volume}{1}, \bibinfo{publisher}{Birckbeck College},
  \bibinfo{address}{London}, \bibinfo{year}{1961}, pp.
  \bibinfo{pages}{105--123}.
\bibitem[{Cox(1962)}]{cox1962further}
\bibinfo{author}{D.R. Cox}, \bibinfo{title}{Further results on tests of
  separate families of hypotheses}, \bibinfo{journal}{J. Roy. Stat. Soc. B}
  \bibinfo{volume}{24} (\bibinfo{year}{1962}) \bibinfo{pages}{406--424}.
\bibitem[{Cribari-Neto et~al.(2002)Cribari-Neto, Frery and
  Silva}]{cribari2002improved}
\bibinfo{author}{F.~Cribari-Neto}, \bibinfo{author}{A.C. Frery},
  \bibinfo{author}{M.F. Silva}, \bibinfo{title}{Improved estimation of clutter
  properties in speckled imagery}, \bibinfo{journal}{Comput. Stat. Data Anal.}
  \bibinfo{volume}{40} (\bibinfo{year}{2002}) \bibinfo{pages}{801--824}.
\bibitem[{Cysneiros et~al.(2008)Cysneiros, Cribari-Neto and
  Ara{\'u}jo}]{cysneiros2008birnbaum}
\bibinfo{author}{A.H. Cysneiros}, \bibinfo{author}{F.~Cribari-Neto},
  \bibinfo{author}{C.A. Ara{\'u}jo}, \bibinfo{title}{On {B}irnbaum-{S}aunders
  inference}, \bibinfo{journal}{Comput. Stat. Data Anal.} \bibinfo{volume}{52}
  (\bibinfo{year}{2008}) \bibinfo{pages}{4939--4950}.
\bibitem[{Davison and Hinkley(1997)}]{davison1997bootstrap}
\bibinfo{author}{A.C. Davison}, \bibinfo{author}{D.V. Hinkley},
  \bibinfo{title}{Bootstrap Methods and Their Application},
  \bibinfo{publisher}{Cambridge University Press}, \bibinfo{address}{New York},
  \bibinfo{year}{1997}.
\bibitem[{D\'{\i}az-Garc\'{\i}a and Dom{\i}nguez-Molina(2006)}]{diaz2006some}
\bibinfo{author}{J.A. D\'{\i}az-Garc\'{\i}a}, \bibinfo{author}{J.R.
  Dom{\i}nguez-Molina}, \bibinfo{title}{Some generalisations of
  {B}irnbaum-{S}aunders and sinh-normal distributions}, \bibinfo{journal}{Int.
  Math. Forum} \bibinfo{volume}{1} (\bibinfo{year}{2006})
  \bibinfo{pages}{1709--1727}.
\bibitem[{D\'{\i}az-Garc\'{\i}a and Leiva(2005)}]{diaz2005new}
\bibinfo{author}{J.A. D\'{\i}az-Garc\'{\i}a}, \bibinfo{author}{V.~Leiva},
  \bibinfo{title}{A new family of life distributions based on the elliptically
  contoured distributions}, \bibinfo{journal}{J. Stat. Plan. Infer.}
  \bibinfo{volume}{128} (\bibinfo{year}{2005}) \bibinfo{pages}{445--457}.
\bibitem[{DiCiccio and Martin(1993)}]{diciccio1993simple}
\bibinfo{author}{T.J. DiCiccio}, \bibinfo{author}{M.A. Martin},
  \bibinfo{title}{Simple modifications for signed roots of likelihood ratio
  statistics}, \bibinfo{journal}{J. Roy. Stat. Soc. B} \bibinfo{volume}{55}
  (\bibinfo{year}{1993}) \bibinfo{pages}{305--316}.
\bibitem[{Doornik(2009)}]{doornik2009object}
\bibinfo{author}{J.A. Doornik}, \bibinfo{title}{An Object-Oriented Matrix
  Programming Language Ox 6.}, \bibinfo{publisher}{Timberlake Consultants
  Press}, \bibinfo{address}{London}, \bibinfo{year}{2009}.
\bibitem[{Efron(1990)}]{efron1990more}
\bibinfo{author}{B.~Efron}, \bibinfo{title}{More efficient bootstrap
  computations}, \bibinfo{journal}{J. Am. Stat. Assoc.} \bibinfo{volume}{85}
  (\bibinfo{year}{1990}) \bibinfo{pages}{79--89}.
\bibitem[{Ehsani et~al.(1996)Ehsani, Saadatmanesh and Tao}]{ehsani1996design}
\bibinfo{author}{M.R. Ehsani}, \bibinfo{author}{H.~Saadatmanesh},
  \bibinfo{author}{S.~Tao}, \bibinfo{title}{Design recommendations for bond of
  {GFRP} rebars to concrete}, \bibinfo{journal}{J. Struct. Eng.}
  \bibinfo{volume}{122} (\bibinfo{year}{1996}) \bibinfo{pages}{247--254}.
\bibitem[{Ferrari and Pinheiro(2016)}]{ferrari2014small}
\bibinfo{author}{S.L. Ferrari}, \bibinfo{author}{E.C. Pinheiro},
  \bibinfo{title}{Small-sample one-sided testing in extreme value regression
  models}, \bibinfo{journal}{Adv. Stat. Anal.} \bibinfo{volume}{100}
  (\bibinfo{year}{2016}) \bibinfo{pages}{79--97}.
\bibitem[{Firth(1993)}]{firth1993bias}
\bibinfo{author}{D.~Firth}, \bibinfo{title}{Bias reduction of maximum
  likelihood estimates}, \bibinfo{journal}{Biometrika} \bibinfo{volume}{80}
  (\bibinfo{year}{1993}) \bibinfo{pages}{27--38}.
\bibitem[{Folks and Chhikara(1978)}]{folks1978inverse}
\bibinfo{author}{J.~Folks}, \bibinfo{author}{R.~Chhikara}, \bibinfo{title}{The
  inverse {G}aussian distribution and its statistical application--a review},
  \bibinfo{journal}{J. Roy. Stat. Soc. B} \bibinfo{volume}{40}
  (\bibinfo{year}{1978}) \bibinfo{pages}{263--289}.
\bibitem[{Fraser et~al.(1999)Fraser, Reid and Wu}]{fraser1999simple}
\bibinfo{author}{D.A.S. Fraser}, \bibinfo{author}{N.~Reid},
  \bibinfo{author}{J.~Wu}, \bibinfo{title}{A simple general formula for tail
  probabilities for frequentist and {B}ayesian inference},
  \bibinfo{journal}{Biometrika} \bibinfo{volume}{86} (\bibinfo{year}{1999})
  \bibinfo{pages}{249--264}.
\bibitem[{G{\'o}mez et~al.(2011)G{\'o}mez, Elal-Olivero, Salinas and
  Bolfarine}]{gomez2011bimodal}
\bibinfo{author}{H.W. G{\'o}mez}, \bibinfo{author}{D.~Elal-Olivero},
  \bibinfo{author}{H.S. Salinas}, \bibinfo{author}{H.~Bolfarine},
  \bibinfo{title}{Bimodal extension based on the skew-normal distribution with
  application to pollen data}, \bibinfo{journal}{Environmetrics}
  \bibinfo{volume}{22} (\bibinfo{year}{2011}) \bibinfo{pages}{50--62}.
\bibitem[{Leiva(2015)}]{leiva2015birnbaum}
\bibinfo{author}{V.~Leiva}, \bibinfo{title}{The Birnbaum-Saunders
  Distribution}, \bibinfo{publisher}{Academic Press},
  \bibinfo{address}{London}, \bibinfo{year}{2015}.
\bibitem[{Leiva et~al.(2010)Leiva, Sanhueza, Kotz and
  Araneda}]{leiva2010unified}
\bibinfo{author}{V.~Leiva}, \bibinfo{author}{A.~Sanhueza},
  \bibinfo{author}{S.~Kotz}, \bibinfo{author}{N.~Araneda}, \bibinfo{title}{A
  unified mixture model based on the inverse {G}aussian distribution},
  \bibinfo{journal}{Pak. J. Stat.} \bibinfo{volume}{26} (\bibinfo{year}{2010})
  \bibinfo{pages}{445--460}.
\bibitem[{Leiva et~al.(2015)Leiva, Tejo, Guiraud, Schmachtenberg, Orio and
  Marmolejo-Ramos}]{leiva2015modeling}
\bibinfo{author}{V.~Leiva}, \bibinfo{author}{M.~Tejo},
  \bibinfo{author}{P.~Guiraud}, \bibinfo{author}{O.~Schmachtenberg},
  \bibinfo{author}{P.~Orio}, \bibinfo{author}{F.~Marmolejo-Ramos},
  \bibinfo{title}{Modeling neural activity with cumulative damage
  distributions}, \bibinfo{journal}{Biol. Cybern.} \bibinfo{volume}{109}
  (\bibinfo{year}{2015}) \bibinfo{pages}{421--433}.
\bibitem[{Lemonte(2016)}]{Lemonte_2016}
\bibinfo{author}{A.J. Lemonte}, \bibinfo{title}{The Gradient Test: Another
  Likelihood-Based Test}, \bibinfo{publisher}{Academic Press},
  \bibinfo{address}{New York}, \bibinfo{year}{2016}.
\bibitem[{Lemonte et~al.(2007)Lemonte, Cribari-Neto and
  Vasconcellos}]{lemonte2007improved}
\bibinfo{author}{A.J. Lemonte}, \bibinfo{author}{F.~Cribari-Neto},
  \bibinfo{author}{K.L.P. Vasconcellos}, \bibinfo{title}{Improved statistical
  inference for the two-parameter {B}irnbaum--{S}aunders distribution},
  \bibinfo{journal}{Comput. Stat. Data Anal.} \bibinfo{volume}{51}
  (\bibinfo{year}{2007}) \bibinfo{pages}{4656--4681}.
\bibitem[{Lemonte and Ferrari(2011)}]{lemonte2011signed}
\bibinfo{author}{A.J. Lemonte}, \bibinfo{author}{S.L. Ferrari},
  \bibinfo{title}{Signed likelihood ratio tests in the {B}irnbaum--{S}aunders
  regression model}, \bibinfo{journal}{J. Stat. Plan. Infer.}
  \bibinfo{volume}{141} (\bibinfo{year}{2011}) \bibinfo{pages}{1031--1040}.
\bibitem[{Lemonte et~al.(2008)Lemonte, Simas and
  Cribari-Neto}]{lemonte2008bootstrap}
\bibinfo{author}{A.J. Lemonte}, \bibinfo{author}{A.B. Simas},
  \bibinfo{author}{F.~Cribari-Neto}, \bibinfo{title}{Bootstrap-based improved
  estimators for the two-parameter {B}irnbaum--{S}aunders distribution},
  \bibinfo{journal}{J. Stat. Comput. Simul.} \bibinfo{volume}{78}
  (\bibinfo{year}{2008}) \bibinfo{pages}{37--49}.
\bibitem[{Lewis et~al.(2011)Lewis, Butler and Gilbert}]{lewis2011unified}
\bibinfo{author}{F.~Lewis}, \bibinfo{author}{A.~Butler},
  \bibinfo{author}{L.~Gilbert}, \bibinfo{title}{A unified approach to model
  selection using the likelihood ratio test}, \bibinfo{journal}{Methods Ecol.
  Evol.} \bibinfo{volume}{2} (\bibinfo{year}{2011}) \bibinfo{pages}{155--162}.
\bibitem[{Liseo(1990)}]{liseo1990classe}
\bibinfo{author}{B.~Liseo}, \bibinfo{title}{La classe delle densita normali
  sghembe: aspetti inferenziali da un punto di vista {B}ayesiano},
  \bibinfo{journal}{Statistica} \bibinfo{volume}{50} (\bibinfo{year}{1990})
  \bibinfo{pages}{71--79}.
\bibitem[{Mittelhammer et~al.(2000)Mittelhammer, Judge and
  Miller}]{mittelhammer2000econometric}
\bibinfo{author}{R.C. Mittelhammer}, \bibinfo{author}{G.G. Judge},
  \bibinfo{author}{D.J. Miller}, \bibinfo{title}{Econometric Foundations},
  \bibinfo{publisher}{Cambridge University Press}, \bibinfo{address}{New York},
  \bibinfo{year}{2000}.
\bibitem[{Ng et~al.(2003)Ng, Kundu and Balakrishnan}]{ng2003modified}
\bibinfo{author}{H.~Ng}, \bibinfo{author}{D.~Kundu},
  \bibinfo{author}{N.~Balakrishnan}, \bibinfo{title}{Modified moment estimation
  for the two-parameter {B}irnbaum--{S}aunders distribution},
  \bibinfo{journal}{Comput. Stat. Data Anal.} \bibinfo{volume}{43}
  (\bibinfo{year}{2003}) \bibinfo{pages}{283--298}.
\bibitem[{Olmos et~al.(2017)Olmos, Martinez-Florez and Bolfarine}]{olmos2015}
\bibinfo{author}{N.~Olmos}, \bibinfo{author}{G.~Martinez-Florez},
  \bibinfo{author}{H.~Bolfarine}, \bibinfo{title}{Bimodal {B}irnbaum-{S}aunders
  distribution with applications to non-negative measurements},
  \bibinfo{journal}{Commun. Stat.-Theor. Methods} \bibinfo{volume}{46}
  (\bibinfo{year}{2017}) \bibinfo{pages}{6240--6257}.
\bibitem[{Owen(2006)}]{owen2006new}
\bibinfo{author}{W.J. Owen}, \bibinfo{title}{A new three-parameter extension to
  the {B}irnbaum--{S}aunders distribution}, \bibinfo{journal}{IEEE Trans.
  Reliab.} \bibinfo{volume}{55} (\bibinfo{year}{2006})
  \bibinfo{pages}{475--479}.
\bibitem[{Owen and Ng(2015)}]{owen2015revisit}
\bibinfo{author}{W.J. Owen}, \bibinfo{author}{H.K.T. Ng},
  \bibinfo{title}{Revisit of relationships and models for the
  {B}irnbaum--{S}aunders and inverse-{G}aussian distributions},
  \bibinfo{journal}{J. Stat. Distrib. Appl.} \bibinfo{volume}{2}
  (\bibinfo{year}{2015}) \bibinfo{pages}{1--23}.
\bibitem[{Patriota(2012)}]{patriota2012scale}
\bibinfo{author}{A.G. Patriota}, \bibinfo{title}{On scale-mixture
  {B}irnbaum--{S}aunders distributions}, \bibinfo{journal}{J. Stat. Plan.
  Infer.} \bibinfo{volume}{142} (\bibinfo{year}{2012})
  \bibinfo{pages}{2221--2226}.
\bibitem[{Pianto and Cribari-Neto(2011)}]{pianto2011dealing}
\bibinfo{author}{D.M. Pianto}, \bibinfo{author}{F.~Cribari-Neto},
  \bibinfo{title}{Dealing with monotone likelihood in a model for speckled
  data}, \bibinfo{journal}{Comput. Stat. Data Anal.} \bibinfo{volume}{55}
  (\bibinfo{year}{2011}) \bibinfo{pages}{1394--1409}.
\bibitem[{Rocke(1989)}]{rocke1989bootstrap}
\bibinfo{author}{D.M. Rocke}, \bibinfo{title}{Bootstrap {B}artlett adjustment
  in seemingly unrelated regression}, \bibinfo{journal}{J. Am. Stat. Assoc.}
  \bibinfo{volume}{84} (\bibinfo{year}{1989}) \bibinfo{pages}{598--601}.
\bibitem[{Sanhueza et~al.(2008)Sanhueza, Leiva and
  Balakrishnan}]{sanhueza2008generalized}
\bibinfo{author}{A.~Sanhueza}, \bibinfo{author}{V.~Leiva},
  \bibinfo{author}{N.~Balakrishnan}, \bibinfo{title}{The generalized
  {B}irnbaum--{S}aunders distribution and its theory, methodology, and
  application}, \bibinfo{journal}{Commun. Stat.-Theor. Methods}
  \bibinfo{volume}{37} (\bibinfo{year}{2008}) \bibinfo{pages}{645--670}.
\bibitem[{Sartori(2006)}]{sartori2006bias}
\bibinfo{author}{N.~Sartori}, \bibinfo{title}{Bias prevention of maximum
  likelihood estimates for scalar skew normal and skew t distributions},
  \bibinfo{journal}{J. Stat. Plan. Infer.} \bibinfo{volume}{136}
  (\bibinfo{year}{2006}) \bibinfo{pages}{4259--4275}.
\bibitem[{Severini(1999)}]{severini1999empirical}
\bibinfo{author}{T.A. Severini}, \bibinfo{title}{An empirical adjustment to the
  likelihood ratio statistic}, \bibinfo{journal}{Biometrika}
  \bibinfo{volume}{86} (\bibinfo{year}{1999}) \bibinfo{pages}{235--247}.
\bibitem[{Severini(2000)}]{severini2000likelihood}
\bibinfo{author}{T.A. Severini}, \bibinfo{title}{Likelihood {M}ethods in
  {S}tatistics}, \bibinfo{publisher}{Oxford University Press},
  \bibinfo{address}{New York}, \bibinfo{year}{2000}.
\bibitem[{Smith et~al.(2015)Smith, Wang, Wong and Zhou}]{smith2015penalized}
\bibinfo{author}{B.~Smith}, \bibinfo{author}{S.~Wang},
  \bibinfo{author}{A.~Wong}, \bibinfo{author}{X.~Zhou}, \bibinfo{title}{A
  penalized likelihood approach to parameter estimation with integral
  reliability constraints}, \bibinfo{journal}{Entropy} \bibinfo{volume}{17}
  (\bibinfo{year}{2015}) \bibinfo{pages}{4040--4063}.
\bibitem[{Vuong(1989)}]{vuong1989likelihood}
\bibinfo{author}{Q.H. Vuong}, \bibinfo{title}{Likelihood ratio tests for model
  selection and non-nested hypotheses}, \bibinfo{journal}{Econometrica}
  \bibinfo{volume}{57} (\bibinfo{year}{1989}) \bibinfo{pages}{307--333}.
\bibitem[{Williams(1970)}]{williams1970discrimination}
\bibinfo{author}{D.A. Williams}, \bibinfo{title}{Discrimination between
  regression models to determine the pattern of enzyme synthesis in synchronous
  cell cultures}, \bibinfo{journal}{Biometrics} \bibinfo{volume}{26}
  (\bibinfo{year}{1970}) \bibinfo{pages}{23--32}.
\bibitem[{Wu and Wong(2004)}]{wu2004improved}
\bibinfo{author}{J.~Wu}, \bibinfo{author}{A.C. Wong}, \bibinfo{title}{Improved
  interval estimation for the two-parameter {B}irnbaum--{S}aunders
  distribution}, \bibinfo{journal}{Comput. Stat. Data Anal.}
  \bibinfo{volume}{47} (\bibinfo{year}{2004}) \bibinfo{pages}{809--821}.
\bibitem[{Zhu and Balakrishnan(2015)}]{zhu2015birnbaum}
\bibinfo{author}{X.~Zhu}, \bibinfo{author}{N.~Balakrishnan},
  \bibinfo{title}{{B}irnbaum--{S}aunders distribution based on {L}aplace kernel
  and some properties and inferential issues}, \bibinfo{journal}{Stat. Probab.
  Lett.} \bibinfo{volume}{101} (\bibinfo{year}{2015}) \bibinfo{pages}{1--10}.

\end{thebibliography}

\end{document}